\documentclass[a4paper]{JHEP3}

\usepackage{amsmath}
\usepackage{latexsym}
\usepackage{cite}
\usepackage{graphicx}

\newcommand{\gthree}{{\gamma^3}}

\newcommand{\medsp}{\\[0.7ex]}

\newcommand{\diff}[1][]{\mbox{d}#1}

\newcommand{\half}[1]{\ensuremath{\frac{#1}{2}}}
\newcommand{\intd}[1]{\int \!\! #1 \;}
\newcommand{\inv}[1]{\ensuremath{\frac{1}{#1}}}

\newcommand{\Stext}[1]{\itindex{\mathcal{S}}{#1}}

\newcommand{\derfrac}[2][]{\frac{\partial #1}{\partial #2}}

\newcommand{\itindex}[2]{\ensuremath{#1_{\mbox{\scriptsize{\itshape #2}}}}}

\newcommand{\varfrac}[2][]{\frac{\delta #1}{\delta #2}}

\DeclareMathOperator{\extdm}{d}
\newcommand{\extd}{\extdm \!}

\title{Graded Poisson-Sigma Models and Dilaton-Deformed
2D Supergravity Algebra}
\author{Luzi Bergamin  and Wolfgang Kummer\\ Institute for Theoretical Physics, Technical University of
  Vienna, Wiedner Hauptstr.\ 8-10, 1040 Vienna, Austria\\ E-mail:
  \email{bergamin@tph.tuwien.ac.at}, \email{wkummer@tph.tuwien.ac.at}}
\abstract{Fermionic extensions of generic 2d gravity theories obtained from the
graded Poisson-Sigma model (gPSM) approach show a large degree of
ambiguity. In addition, obstructions may reduce the allowed range of fields as
given by the bosonic theory, or even prohibit any extension in certain
cases. In our present work we relate the finite W-algebras inherent
in the gPSM algebra of constraints to algebras which can be interpreted as
supergravities in the usual sense (Neuveu-Schwarz or Ramond algebras resp.),
deformed by the presence of the dilaton field.
With very straightforward and natural assumptions on them --like demanding rigid supersymmetry in
a certain flat limit, or  linking the anti-commutator of certain fermionic charges to the
Hamiltonian constraint-- in the ``genuine'' supergravity obtained in this way
the ambiguities disappear, as well as the obstructions referred to above.
Thus all especially interesting bosonic models (spherically reduced
gravity, the Jackiw-Teitelboim model etc.)\ under these
conditions possess a unique fermionic extension and are free from
new singularities. The superspace supergravity model of Howe is found as
a special case of this supergravity action. For this class of models the relation between bosonic
potential and prepotential does not introduce obstructions as well.}
\preprint{hep-th/0209209\\TUW-02-09}
\keywords{Dilaton supergravity, graded Poisson-Sigma model, superconformal algebra}

\begin{document}

\bibliographystyle{JHEP}

\section{Introduction}
\label{sec:introd}
Diffeomorphism invariant dilaton theories in 1+1 dimensions for some
time have been a promising field in classical and quantum theory.
They include effective theories of direct physical interest, like
reduced d-dimensional Einstein theories and the extensions thereof
(Einstein-deSitter, Jordan-Brans-Dicke theories \cite{Dicke:1957,jordanschwerkraft,Jordan:1959eg,Fierz:1956,Brans:1961sx}), but
also theories suggested by stringy arguments \cite{Callan:1992rs}
\footnote{A recent review with rather extensive literature is represented by ref.\
\cite{Grumiller:2002nm}.}. On the other hand, supersymmetric extensions of gravity
\cite{Freedman:1976xh,Freedman:1976py,Deser:1976eh,Deser:1976rb}
are believed to be a crucial ingredient for a consistent solution
of the problem how to quantize gravity in the framework of string/brane
theory \cite{Green:1987sp,lust89,Polchinski:1998rq}.

Much of the recent progress \cite{Haider:1994cw,Kummer:1997hy,Kummer:1998jj,Kummer:1998zs,Grumiller:2002nm} to understand bosonic gravity
theories in two dimensions also at the quantum level is based upon the
equivalence \cite{Katanaev:1996bh,Katanaev:1997ni} of a 
torsion free general dilaton theory
\cite{Mann:1990gh,Banks:1991mk,Gegenberg:1993rg,Lemos:1994py,Louis-Martinez:1994eh}
and a Hamiltonian action of the type of
a Poisson-Sigma model (PSM)\cite{Schaller:1994np,Schaller:1994es}. A (graded)
PSM ((g)PSM) is defined by the action
\begin{equation}
  \label{eq:action}
  \begin{split}
    \mathcal{S} &= \int_M \extd X^I \wedge A_I + \half{1} P^{IJ} A_J \wedge A_I \medsp
    &= \intd{\diff{^2 x}} \bigl(\partial_0 X^I A_{1 I} - \partial_1 X^I A_{0I} + P^{IJ}
    A_{0J} A_{1I} \bigr)\ .
  \end{split}
\end{equation}

Target space coordinates \( X^{I}(x) \) and one-form gauge-fields $A_I =
A_{mI}(x) \extd x^m$
live upon the 2d manifold $M$. The dynamics is encoded in the Poisson
tensor \( P^{IJ}(X) \), which must have a vanishing (graded) Nijenhuis
tensor\footnote{Here and in the following $\partial_K = \derfrac{X^K}$ represents the
  derivative with respect to the target space variables. More details of the
  conventions are given in the Appendix.}
(obey the Jacobi identity for a (graded) Schouten bracket defined as $[ X^I, X^J ] = P^{IJ}$)
\begin{equation}
\label{eq:nijenhuis}
  P^{IL}\partial _{L}P^{JK}+ \mbox{\it 
g-perm}\left( IJK\right) = 0\ .
\end{equation}
Then the action \eqref{eq:action} is invariant under the symmetry transformations
\begin{align}
\label{eq:symtrans}
  \delta X^{I} &= P^{IJ} \epsilon _{J}\ , & \delta A_{I} &= -\mbox{d} \epsilon
  _{I}-\left( \partial _{I}P^{JK}\right) \epsilon _{K}\, A_{J}\ ,
\end{align}
where the term $\extd \epsilon_I$ in the second of these equations provides
the justification for calling $A_I$ ``gauge fields''.

For a generic (g)PSM the commutator of two transformations \eqref{eq:symtrans} is a
symmetry modulo the equations of motion. Only for \( P^{IJ} \) linear
in \( X^{I} \) a closed (and linear) Lie algebra is obtained and
\eqref{eq:nijenhuis} reduces to the Jacobi identity for the structure
constants of a Lie group. If the Poisson
tensor is singular --the actual situation in any application to 2d
(super-)gravity due to the odd dimension of the bosonic part of the tensor--
there exist (one or more) Casimir functions $C(X)$ obeying
\begin{equation}
\label{eq:casimir}
  [ X^I, C ] = P^{IJ}\derfrac[C]{X^J} = 0\ ,
\end{equation}
which, when determined by the field equations of motion, are constants of
motion.

So far the number of target space coordinates (the range of the indices $I$,
$J$ etc.)\ has not been specified. In the following we identify\footnote{Further
  explanations of the usage of indices are given in the Appendix.} the first
component $X^0$ of $X^I$ with the dilaton field $\phi$ and use the symbol
$\phi$ also to describe that value of $I$: $X^0 = X^\phi = \phi$. The
corresponding gauge field becomes $A_0 = A_\phi = \omega$.
In the application to bosonic gravity\footnote{The condition to identify the PSM 1-forms
  directly with the geometrical variables can be relaxed, leading to more
  general gravity theories. Given the mathematical difficulties to find
  explicit solutions with vanishing Nijenhuis tensor even for the present
  situation, we restrict the discussion to this direct identification.} $A_{I}$
comprises the Cartan variables spin connection \( \omega _{ab}=\omega \epsilon _{ab}  \)
and zweibeine \( e_{a}  \). The
corresponding three target
space coordinates are \( X^{I} = X^i =\left( \phi ,X^{a}\right)  \). The
component \( P^{\phi a}=X^{b}\epsilon _{b}^{\, \, a} \)
is determined by local Lorentz invariance. Only this choice leads to the
correct covariant derivative $(D e)_a = \extd e_a - \omega {\epsilon_a}^b e_b$
in the 2d gravity action (cf.\ eq.\ \eqref{eq:A8}). In the pure bosonic case
$P^{ij} = - P^{ji}$ is anti-symmetric. Therefore, the only remaining
components of the Poisson tensor can be written as \( P^{ab}=v\epsilon^{ab}
\), where the locally Lorentz invariant 
``potential'' \( v=v\left( \phi ,Y\right)  \) describes different
models ( \( Y=X^{a}X_{a}/2 \) ). Evaluating \eqref{eq:action} with that
$P^{\phi a}$ and $P^{ab}$ the action ($\epsilon = \half{1} \epsilon^{ab} e_b \wedge e_a$)
\begin{equation}
  \label{eq:bosonicPSM}
  \Stext{bosonic} = \int_{\mathcal{M}} \bigl( \phi \diff \omega + X^a D e_a +
  \epsilon v \bigr)
\end{equation}
is obtained. The most
interesting models are described by potentials of type
\begin{equation}
  \label{eq:bosonpot}
  v=Y\, Z\left( \phi \right) +V\left( \phi \right)\ ,
\end{equation}
where e.g.\ spherically reduced gravity from d dimensions
 is given by \cite{Thomi:1984na,Hajicek:1984mz,Schmidt:1997mq,Schmidt:1998ih}
 \begin{align}
\label{eq:SRG}
   Z_{SRG}& = -\frac{d-3}{\left( d-2\right) \phi }\ , & V_{SRG} &= -\lambda
   ^{2}\phi ^{\left( d-4\right) /\left( d-2\right) }\ ,
 \end{align}
with the CGHS model \cite{Callan:1992rs} as the formal limit $d \rightarrow
\infty$. The Jackiw-Teitelboim model
\cite{Barbashov:1979bm,Teitelboim:1983ux,teitelboim:1984,jackiw:1984,Jackiw:1985je}
corresponds
to
\begin{align}
\label{eq:JT}
  Z_{JT} &= 0\ , &  V_{JT} &= -\Lambda \phi\ ,
\end{align}
and the bosonic part of the simplest non-trivial 2d supergravity model of Howe
\cite{Howe:1979ia} --after the correct identification of the dilaton with one
of the components of the superfield \cite{Ertl:2000si}-- becomes
\begin{align}
\label{eq:howe}
Z_{H} &=0\ , & V_{H}&=-\frac{1}{2}\phi ^{2}\ .
\end{align}
Potentials of type \eqref{eq:bosonpot} allow the integration of the (single) Casimir function
$C$ in \eqref{eq:casimir}
\begin{align}
\label{eq:bosonicC}
  C&= e^{Q(\phi)} Y + W(\phi) & Q(\phi) &= \int_{\phi_1}^\phi \extd \varphi
  Z(\varphi) & W(\phi) &= \int_{\phi_0}^\phi \extd \varphi e^{Q(\varphi)}
  V(\varphi)\ ,
\end{align}
where e.g., in spherically reduced gravity \eqref{eq:SRG} $C$ on-shell simply is proportional
to the ADM-mass in the Schwarzschild solution.

The auxiliary variables
\( X^{a} \) and the torsion-dependent part of the spin connection
can be eliminated by \emph{algebraic} equations of motion. Then the action
reduces to the familiar generalized dilaton theory
in terms of the dilaton field \( \phi  \) and the metric:
\begin{equation}
\label{eq:GDT}
  \Stext{GDT} = \intd{\diff{^2 x}} e \Bigl(\half{1} R \phi - \half{1} Z
  \partial^m \phi \partial_m \phi + V \Bigr)
\end{equation}
Both formulations are equivalent at the classical \cite{Katanaev:1996bh,Katanaev:1997ni} as
well as at the quantum level
\cite{Kummer:1997hy,Kummer:1998jj,Kummer:1998zs}. But the PSM formulation
possesses crucial technical advantages, another of which is the topic of our
present work.

A generalization of $\Stext{GDT}$ to supergravity exists for a
long time \cite{Howe:1979ia,Park:1993sd}. However, in order to exploit the
advantages of the formulation as a ``Hamiltonian'' one in
\eqref{eq:bosonicPSM}, a supergravity extension of \emph{that} action is
desirable. All gravity models
with $Y$-dependent potentials in the PSM formulation ($Z \neq 0$ in
\eqref{eq:bosonpot}) possess non-vanishing
bosonic torsion which turned out to be an extremely cumbersome feature
in all attempts to formulate directly corresponding 2d supergravities in the
superfield formalism. The conventional constraints for vanishing bosonic
torsion are well studied. Now they have
to be changed in a highly nontrivial manner. As shown in \cite{Ertl:2001sj}
the entire procedure of solving Bianchi identities for superfields with a
large number of nontrivial components must be started from scratch, if the
bosonic torsion does not vanish any more. In
contrast, the PSM extends straightforwardly by the
introduction of \( N \) additional fermionic target space Majorana coordinates
\( \chi ^{\left( i\right) \alpha } \) (``dilatinos'') and corresponding ``gravitinos''
\( \psi ^{\left( i\right) }_{\alpha } \) \(\left(
  i=1,...N\right)  \) in \( X^{I} \), resp.\ \( A_{I} \)
\begin{align}
  X^I &= (X^i, X^{(n) \alpha}) = (\phi, X^a, \chi^{(n) \alpha})\ , & A_I &= (A_i,
  A^{(n)}_{\alpha}) = (\omega, e_a, \psi^{(n)}_{\alpha})
\end{align}
to a gPSM with a graded Poisson tensor \(
P^{IJ}=-\left( -1\right) ^{IJ}P^{JI} \) \cite{Strobl:1999zz,Ertl:2000si}. Thanks to our conventions (cf.\
Appendix) for
summation of adjacent indices in eqs.\
\eqref{eq:action}-\eqref{eq:casimir} all formulas remain unchanged. This generalization directly provides an (\( N,N \))
``supergravity'' theory without auxiliary fields and thus without the
need to impose constraints, to solve Bianchi identities etc. As can be seen
from the second eq.\ in \eqref{eq:symtrans} the solution of the graded
counterpart of that equation contains a supergravity-like transformation
law for the gravitino-field \( \psi _{\alpha } \), deformed by the
dilaton fields.

When a bosonic potential, as exemplified by \eqref{eq:bosonpot}, is given, the
solution of the graded
counterpart of the Jacobi identity \eqref{eq:nijenhuis} provides all possible
fermionic extensions. It turned
out that an \emph{algebraic} general solution\footnote{For degenerate fermionic
  extensions also one or two additional fermionic Casimirs appear beside the
  basic one in eq.\ \eqref{eq:bosonicC}.} of this problem can be given explicitly
\cite{Ertl:2000si}. However, this
solution was found to be far from unique. As in ref.\ \cite{Ertl:2000si}, in
the present work we also concentrate
on $N = (1,1)$ supergravity, which must be realized as a subset of such gPSM
theories with one field $\chi^\alpha$ in a non-degenerate fermionic
extension (rank 2). In this case \cite{Ertl:2000si} the
general solution for a certain bosonic potential
\( v \) depends on no less than five arbitrary
Lorentz covariant functions. In addition,
the new fermionic terms in the algebras and in the ensuing
Lagrangians produced by this method exhibit new singularities (obstructions) in
the variables $\phi$ and $Y$ in points where the original bosonic potential
was regular. For certain bosonic models
\emph{any} fermionic extension is prohibited. The basic reason for the latter
behaviour is that the potential \eqref{eq:bosonpot} in the extended case must
be derivable from a prepotential.

Attempts to simply require the absence of new singularities in the fermionic
part of the action as a means to determine {}``allowed'' extensions
turned out to be unsuccessful: In a counter-example \cite{Ertl:2000si} two
\textit{different}
supergravities with the \emph{same} (bosonic) \( v \) of spherical
reduction (\eqref{eq:bosonpot} with \eqref{eq:SRG}) were found to be free from such singularities.

Therefore, the question arises how the additional structure present in
\emph{standard} supergravity theories formulated in superspace can be found within
gPSM gravity models and whether the unpleasant features of a generic gPSM
described above are influenced by the ensuing restrictions.

Obviously the nonlinear gauge symmetry \eqref{eq:symtrans} does not lend
itself to an appropriate
starting point for this task, the conditions for supergravity being not inherent in
general gPSM theories: Any strategy must --in some sense-- contain the
restriction known in supergravity models from superspace or from a
gauge-theoretic approach that, in the limit of flat space-time of the bosonic
geometry, the fermionic sector must reduce to rigid supersymmetry
\cite{Freedman:1976xh,Freedman:1976py,Deser:1976eh,Deser:1976rb,Grimm:1978kp}.
In a generic PSM the bosonic potential need not have any flat space-time limit
and thus a generalization covering also those geometries is
necessary. To implement a condition, which is as close as possible to this
basic principle, we propose to start from the closed but
nonlinear algebra of Hamiltonian
constraints $G^I = (G^\phi, G^a, G^\alpha)$, obtained for bosonic PSM-models in refs.\
\cite{Grosse:1992vc,Schaller:1994np} and consider their straightforward
generalizations to gPSMs. However, instead of trying to impose supergravity-type
conditions on that nonlinear algebra, we map it onto a linear Virasoro-like algebra
deformed by the presence of the dilaton field. This step again is motivated from the
study of purely bosonic PSM gravities: 
Proper linear combinations of the PSM constraints $G^i$ derived from
$\Stext{bosonic}$ \eqref{eq:bosonicPSM} are known
to become the ADM constraints \( H_{\left( 0\right) } \) and \( H_{\left( 1\right) } \)
\cite{Katanaev:1994qf,Katanaev:2000kc}, which are related to lapse and shift
in the ADM formulation \cite{Arnowitt:1962} of gravity. Together with the Lorentz constraint
$G^\phi$ they form a \emph{linear} (Virasoro-type) algebra.

The strategy of our paper is to
extend \emph{that} algebra to its graded version derived from a gPSM. In contrast to
the bosonic models, this superalgebra will turn out to be extremely
complicated. It is neither linear nor does it represent
a drastic simplification compared to the algebra of constraints $G^I$. In particular
it is not obvious to interpret the general result as a deformed Neuveu-Schwarz
or Ramond algebra\footnote{In the following, for simplicity, we shall use the term
  ``superconformal'' algebra, in agreement the nomenclature used in string theory.}. It rather
reflects in a cumbersome way the large arbitrariness in the gPSM as derived by the fermionic
extension from a PSM.

Clearly the study of \emph{rigid} supersymmetry in flat
space in the gPSM framework is a necessary first step. It permits the
identification of the proper fermionic constraints $I_{(\pm)}$ from $G^\alpha$, to be used beside
the bosonic ones ($H_{(0)}$, $H_{(1)}$, $G^\phi$ above). On this basis three restrictions appear
to be ``natural''. We propose
for ``genuine'' deformed supergravities that

\renewcommand{\labelenumi}{(\arabic{enumi})}
\begin{enumerate}
\item rigid supersymmetry appears, when the deformation by the dilaton field is removed in flat
space.
\item the mathematical structure of $H_{(0)}$, $H_{(1)}$ and of the new fermionic
  generators $I_{(\pm)}$ in terms of the Hamiltonian constraints $G^I$ (cf.\
  eqs.\ \eqref{eq:susyham} and \eqref{eq:rigidI} below) is the
  same as in rigid supersymmetry. In particular we do not allow additional
  terms in the general form of the generators, whose prefactors vanish in
  the limit of rigid supersymmetry.
\item the (anti-commuting) Poisson brackets of the constraints  $I_{(+)}$ with
  itself and $I_{(-)}$ with itself lead
back to \( H_{\left( 0\right) } \) and \( H_{\left( 1\right) } \)
and, at most, the Lorentz constraint $G^\phi = G$.
\end{enumerate}
Actually a more precise mathematical formulation will be given in
Section \ref{sec:threetwo} (requirement (2)) and Section \ref{sec:genSUGRA}
(requirement (3)) after the structure of the general algebra has been set
out in detail.

Condition (1) obviously must be fulfilled for any model involving
supergravity. The same holds for the first part of (3), which we,
though, at first weaken by allowing the appearance of
the Lorentz constraint. As will be seen below, however, the conditions
(1) and (2) imply that such a dependence does not occur.
Requirement (2) implies that the simplest form of $H_{(0)}$,
$H_{(1)}$ and $I_{(\pm)}$ compatible with requirement (1) is used.

The very gratifying consequence of (1) to (3), the central
result of our paper, is that these requirements not only lead
to a nontrivial result, but that the generic problems of the most general
supergravities from gPSMs \cite{Ertl:2000si,Ertl:2001sj} disappear in the
particular class of models selected by those requirements.
Moreover that class precisely covers the physically interesting models
whose bosonic content is given by \eqref{eq:bosonpot}-\eqref{eq:howe}.

The paper is organized as follows: In Section \ref{sec:constralg} we present the algebra of Hamiltonian constraints for
a generic gPSM and review the introduction of the ADM constraints
for the pure bosonic case. Then the gPSM-formalism is applied to superalgebras
containing rigid supersymmetry (Section \ref{sec:gengPSM}). The three restrictions,
set out above, are tested in Section \ref{sec:dilatonprepot} for the model of ref.\
\cite{Izquierdo:1998hg} without bosonic torsion (\( Z=0 \) in \eqref{eq:bosonpot}), which covers the model
of refs.\
\cite{Barbashov:1979bm,Teitelboim:1983ux,teitelboim:1984,jackiw:1984,Jackiw:1985je}
and ref.\ \cite{Howe:1979ia} too (cf.\ \eqref{eq:JT} and \eqref{eq:howe}). 

Section \ref{sec:genSUGRA} is devoted to the analysis of the consequences in the general
case, the final form of the superalgebra and the only allowed class of
``genuine'' supergravity actions as derived from gPSMs are presented in Section
\ref{sec:finalalg}. In the Conclusion (Sect.\ \ref{sec:conclusions}) we
summarize the results. The Appendix contains a summary of our notations.


\section{Covariant algebra in (graded) PSM}
\label{sec:constralg}
\subsection{(g)PSM Hamiltonian constraints}
The constraints $G^I$ for gPSMs are the result of 
a Hamiltonian analysis of the action \eqref{eq:action}.
Canonical variables\footnote{The somewhat unusual
  association of gauge fields as ``momenta''
and of target space coordinate fields as ``coordinates'' is appropriate
when natural boundary conditions (\( \delta X^{I}=0 \) at \( \partial M \))
are assumed. Also quantum ordering problems are eliminated in this
way \cite{Bergamin:2001}.} and first class primary
constraints are defined from the Lagrangian $L$ in \eqref{eq:action}
($\dot{q}^I = \partial_0 X^I$) by
\begin{align}
  X^I &= q^I\ , & \bar{q}^I & \approx 0\ , \medsp
\label{eq:gaugepot}
   \derfrac[L]{\dot{q}^I} &= p_I = A_{1 I}\ , & \bar{p}_I &= A_{0 I}\ .
\end{align}
From the Hamiltonian density ($\partial_1 = \partial$)
\begin{equation}
\label{eq:hamdensity}
  H =  \dot{q}^I p_I - L = \partial q^I \bar{p}_I - P^{IJ} \bar{p}_J p_I
\end{equation}
the graded canonical equations
\begin{align}
  \derfrac[H]{p_I} &= (-1)^I \dot{q}^I\ , & \derfrac[H]{q^I} &= - \dot{p}_I
\end{align}
are consistent with the graded Poisson\footnote{This is the standard Poisson
  bracket, not to be confused with the Schouten bracket, associated to the
  Poisson tensor $P^{IJ}$ in eqs.\ \eqref{eq:nijenhuis}, \eqref{eq:symtrans}.}
  bracket for functionals $A$ and $B$
\begin{equation}
\label{eq:canonicalbr}
  \begin{split}
    \{A, B'\} &= \int_{x''} \Bigl[\bigl( (-1)^{A \cdot I} \varfrac[A]{{q''}^I}
    \varfrac[B']{p''_I} - (-1)^{I(A+1)} \varfrac[A]{p''_I}
    \varfrac[B']{{q''}^I}\bigr) + (q\rightarrow \bar{q}, p \rightarrow
    \bar{p}) \Bigr]\medsp
    &= \int_{x''} \biggl[ \Bigl(\bigl( \varfrac[A]{{q''}^i}
    \varfrac[B']{p''_i} - \varfrac[A]{p''_i}
    \varfrac[B']{{q''}^i}\bigl) + (-1)^A \bigl(  \varfrac[A]{{q''}^\alpha}
    \varfrac[B']{p''_\alpha} + \varfrac[A]{p''_\alpha}
    \varfrac[B']{{q''}^\alpha}\bigr)\Bigr)\medsp
    &\phantom{= \int_{x''} \biggl[ \Bigl(\bigl(}+ (q\rightarrow \bar{q}, p \rightarrow
    \bar{p}) \biggr]\ ,
  \end{split}
\end{equation}
where $(q\rightarrow \bar{q}, p \rightarrow \bar{p})$ indicates that the
functional derivatives have to be performed for both types of variables, with and without bar.
The primes indicate the dependence on primed world-sheet
coordinates $x$, resp.\ $x'$, $x''$. The Hamiltonian density \eqref{eq:hamdensity}
\begin{equation}
  H = G^I \bar{p}_I
\end{equation}
is expressed in terms of secondary constraints only:
\begin{equation}
  \{\bar{q}^I, \intd{\diff{x^1}} H \} = G^I = \partial q^I + P^{IJ} p_J 
\end{equation}

The complete supersymmetric algebra of the gPSM after a straightforward but
tedious calculation with the graded Poisson bracket \eqref{eq:canonicalbr} takes the simple form ($\partial \delta(x' -  x) =
\partial_{x^1} \delta(x^1 - {x'}^1)$)
\begin{align}
\begin{split}
\label{eq:canonicalalg}
  \{ q^I, p'_J \} &= (-1)^I \delta(x - x') \delta^I_J\ ,\\
  \{ \bar{q}^I, \bar{p}'_J \} &= (-1)^I \delta(x - x') \delta^I_J\ , \\
  \{ q^I, \bar{p}'_J \} &= \{ \bar{q}^I, p'_J \} = 0\ , \end{split} \medsp
  \label{eq:constralg} \{ G^I, {G'}^J \} &= - G^K \partial_K P^{IJ} \delta(x -
  x')\ ,
  \medsp
  \begin{split}\label{eq:canonicalalgII} \{ G^I, \bar{p}'_J \} &= \{ G^I, {\bar{q}'}{^J} \} = 0\ ,\\
  \{ G^I, p'_J \} &= (-1)^I \partial \delta(x - x') \delta^I_J + (-1)^{IJ}
  \partial_J P^{IK} p_K \delta(x - x')\ , \\
  \{ G^I, {q'}^J \} &= -P^{IJ} \delta(x-x')\ ,
\end{split}
\end{align}
which is seen to preserve the structure of the bosonic case \cite{Grosse:1992vc}.
Eq.\ \eqref{eq:constralg} implies that the constraint algebra in
general is a finite $W$-algebra with structure functions ${f_K}^{IJ} =
\partial_K P^{IJ}$. It simplifies to a Lie-algebra if the Poisson tensor is
linear in the target-space coordinates, only.
\subsection{Conformal algebra for PSMs}
It is a remarkable
result for non-supersymmetric models that such a $W$-algebra in the absence of
fermionic fields can be mapped
onto the standard current-algebra of conformal symmetry (Virasoro algebra). In ref.\
\cite{Katanaev:1994qf} for two
dimensional gravity with (the anholonomic index $a$ is expressed in light-cone
components, cf.\ Appendix) 
\begin{align}
  q^i &= (q^\phi, q^a) =  (\phi, X^{++}, X^{--})\ , && \medsp
  p_i &= (p_\phi, p_a) = ( \omega_1, e_{1|++}, e_{1|--})\ , & \bar{p}_i
  &= (\bar{p}_\phi, \bar{p}_a) = (\omega_0, e_{0|++}, e_{0|--} )\ ,
\end{align}
from an ADM analysis
\cite{Arnowitt:1962} a different form of the Hamiltonian
density $H = N_{(0)}  H_{(0)}
+ N_{(1)} H_{(1)} + B_{(0)} G$ was obtained. Its constraints become Lorentz invariant linear combinations of the $G^i = (G^\phi, G^a) = (G^\phi,
G^{++}, G^{--})$:
\begin{align}
\label{eq:admham}
  H_{(0)} &= - G^a {\epsilon_a}^b p_b =  G^{++} p_{++} - G^{--} p_{--} &
  H_{(1)} &= G^a p_a = G^{++} p_{++} + G^{--} p_{--}
\end{align}
Here $H_{(0)}$ and $H_{(1)}$ are related to the lapse $N_{(0)}$ and shift
$N_{(1)}$, respectively. Because of \eqref{eq:canonicalalg}-\eqref{eq:canonicalalgII} they
generate another algebra (``deformed ADM algebra'') with the
Lorentz constraint $G^\phi = G$ mixed in:
\begin{align}
  \begin{split}
\label{eq:virasoro2.1}
    \{ H_{(0)}, H'_{(0)} \} &=  (H_{(1)} + H'_{(1)}) \partial
    \delta(x-x') \\
    \{ H_{(0)}, H'_{(1)} \} &=  (H_{(0)} + H'_{(0)}) \partial
    \delta(x-x') + 2 G (\partial_\phi P^{-- k}) p_k p_{--} \delta (x-x')\\
    \{ H_{(1)}, H'_{(1)} \} &=  (H_{(1)} + H'_{(1)}) \partial
    \delta(x-x')
  \end{split} \medsp
\label{eq:virasoro2.2}
    \{G, H'_{(0)} \} &= \{G, H'_{(1)} \} = \{ G, G' \} = 0
\end{align}
By a modification of the ADM constraints \cite{Katanaev:2000kc} in terms of $G
= G^\phi$ and
\begin{equation}
\label{eq:psmham}
  \begin{split}
    \tilde{H}_{(0)} &= G^\phi p_\phi + G^{++} p_{++} - G^{--} p_{--}\ , \\
    \tilde{H}_{(1)} &= G^i p_i\ ,
  \end{split}
\end{equation}
even a \emph{linear} algebra emerges that closes entirely under derivatives of
$\delta$-functions and contains $G$ as a semi-direct product:
\begin{align}
  \begin{split}
\label{eq:virasoro1.1}
    \{ \tilde{H}_{(0)}, \tilde{H}'_{(0)} \} &=  (\tilde{H}_{(1)} + \tilde{H}'_{(1)}) \partial
    \delta(x-x') \\
    \{ \tilde{H}_{(0)}, \tilde{H}'_{(1)} \} &=  (\tilde{H}_{(0)} + \tilde{H}'_{(0)}) \partial
    \delta(x-x') \\
    \{ \tilde{H}_{(1)}, \tilde{H}'_{(1)} \} &=  (\tilde{H}_{(1)} + \tilde{H}'_{(1)}) \partial
    \delta(x-x')
  \end{split} \medsp
  \begin{split}
\label{eq:virasoro1.2}
    \{G, \tilde{H}'_{(0)} \} &= G' \partial \delta(x-x') \\
    \{G, \tilde{H}'_{(1)} \} &= G' \partial \delta(x-x') \\
    \{ G, G' \} &= 0
  \end{split}
\end{align}
By defining the (anti-)holomorphic
 ``currents'' $\tilde{H}_{(\pm)} = \half{1}
(\tilde{H}_{(0)} \pm \tilde{H}_{(1)})$, \eqref{eq:virasoro1.1} transforms into the
standard Virasoro algebra of string theory.
Thus, although the original PSM was formulated in terms of a non-linear
 $W$-algebra for the PSM constraints $G^i$, 
by this simple redefinition the deformation
from the dilaton field is hidden in the non-trivial, but still linear, action of the abelian
Lorentz-constraint $G$.


\section{Algebras for generic gPSM gravity}
\label{sec:gengPSM}
The method of redefining constraints as in \eqref{eq:admham} or
\eqref{eq:psmham} shall now be extended to the graded version of the
PSM, including fermionic generators
$I_{(\pm)}$ besides the ADM constraints $H_{(\pm)}$. 
An essential ingredient for a ``genuine'' supergravity is that
it possesses rigid supersymmetry as a limiting case when the deformations
are turned off and when the metric becomes the one describing flat
Minkowski space. It is, therefore, imperative as a first step to study rigid supersymmetry
in this formalism and then to consider deformations which agree with
the basic principles set out in section \ref{sec:introd}. The choice of
those requirements, of course, will be largely determined by the experience drawn
from this particular case.
\subsection{Rigid supersymmetry}
In the gPSM formulation rigid supersymmetry is produced by the specific choice
of the Poisson tensor \cite{Ertl:2000si}
\begin{gather}
\label{eq:rigidgPSM}
    \begin{alignat}{2}
      P^{a \phi} &= X^b {\epsilon_b}^a\ , &\qquad P^{\alpha \phi} &= -
      \half{1}
      \chi^\beta {\gthree_\beta}^\alpha\ ,
    \end{alignat} \medsp
\label{eq:rigidgPSM2}
    P^{\alpha \beta} = \tilde{u}_0 i X^c \gamma_c^{\alpha \beta} + c
    (\gthree)^{\alpha \beta}\ , \medsp
\label{eq:rigidgPSMlast}
    \begin{alignat}{2}
      P^{ab} &= 0\ , &\qquad P^{a \beta} &= 0\ .
    \end{alignat}
\end{gather}
Here and in the following we restrict by the choice \eqref{eq:rigidgPSM2} to
full $N = (1,1)$ supergravity (``full fermionic rank'' of ref.\ \cite{Ertl:2000si}).
Equations \eqref{eq:rigidgPSM} are uniquely determined by the requirement of Lorentz
covariance (cf.\ \eqref{eq:A8}). The first term in \eqref{eq:rigidgPSM2} is
seen to produce the minimal contribution to the supertorsion, $c$ is an arbitrary constant. Indeed, inserting
\eqref{eq:rigidgPSM}-\eqref{eq:rigidgPSMlast} into \eqref{eq:constralg} yields
\begin{gather}
\label{eq:lorentzconstr}
  \begin{alignat}{2}
    \{ G^a , G^\phi \} &= - G^b {\epsilon_b}^a\ , &\qquad \{ G^\alpha , G^\phi \}
    &= \half{1} G^\beta {\gthree_\beta}^\alpha\ ,
  \end{alignat}\medsp
  \{G^\alpha, G^\beta \} = - \tilde{u}_0 i G^a \gamma_a^{\alpha \beta}\ , \medsp
  \begin{alignat}{2}
    \{ G^a, G^b \} &= 0\ , &\qquad \{ G^a, G^\alpha \} &= 0\ .
  \end{alignat}
\end{gather}
Choosing the normalization $\tilde{u}_0 = 2$ this is the well-known
supersymmetry algebra\footnote{Notice our conventions for the $\gamma$
  matrices in eqs.\ \eqref{eq:gammadef} and \eqref{eq:gammalc}.} with Lorentz
transformations induced by the generator $G^\phi$, translations $G^a$ and supersymmetry transformations
generated by $G^\alpha$.

Analogously to the non-supersymmetric case we now derive the
superconformal algebra by an appropriate choice of new constraints. The
fermionic constraint must be the ``square root'' of the Hamiltonian
constraints, proportional to $G^\alpha = (G^+, G^-)$ and, like $H_{(0)}$ and
$H_{(1)}$, remain invariant under local Lorentz transformations $G^\alpha \rightarrow
(e^\lambda G^+, e^{-\lambda} G^-)$, $p_a \rightarrow (e^{-2 \lambda} p_{++},
e^{2 \lambda} p_{--})$. From these conditions it is easy
to guess that $I_{(+)} \propto G^+ \sqrt{p_{++}}$, $I_{(-)} \propto G^-
\sqrt{p_{--}}$. Indeed, in terms of the constraints of type \eqref{eq:admham}
\begin{align}
\label{eq:susyham}
  H_{(+)} &= G^{++} p_{++} + G^{+} p_{+}\ , &  H_{(-)} &= - G^{--} p_{--} - G^{-}
  p_{-}\ ,
\end{align}
supplemented by the fermionic constraints ($\tilde{u}_0 = 2$)
\begin{align}
\label{eq:rigidI}
  I_{(+)} &= \pm \inv{\sqrt{2 \sqrt{2}}}\; G^+ \sqrt{p_{++}}\ , & I_{(-)} &= \pm
  \inv{\sqrt{2 \sqrt{2}}}\; G^- \sqrt{p_{--}}\ ,
\end{align}
$H_{(\pm)}$ and $I_{(\pm)}$ are found to generate the standard superconformal
algebra, whereas the
Lorentz constraint decouples completely:
  \begin{align}
\label{eq:rigidsusy1}
    \begin{split}
      \{ H_{(+)}, H'_{(+)} \} &=  (H_{(+)} + H'_{(+)}) \partial
    \delta(x-x') \\
      \{ H_{(-)}, H'_{(-)} \} &=  - (H_{(-)} + H'_{(-)}) \partial
    \delta(x-x') \\
      \{ H_{(+)}, H'_{(-)} \} &= 0
    \end{split}\medsp
    \begin{split}\label{eq:rigidsusy2}
      \{ I_{(+)}, I'_{(+)} \} &=  - 2  H_{(+)}  \delta(x-x') \\
      \{ I_{(-)}, I'_{(-)} \} &=  2  H_{(-)}  \delta(x-x') \\
      \{ I_{(+)}, I'_{(-)} \} &=0
    \end{split}\medsp
    \begin{split}
    \{ I_{(+)}, H'_{(+)} \} &=  (I_{(+)} + \half{1} I'_{(+)}) \partial
    \delta(x-x') \\
   \{ I_{(-)}, H'_{(-)} \} &=  - (I_{(-)} + \half{1} I'_{(-)}) \partial
    \delta(x-x') \\
    \{ I_{(-)}, H'_{(+)} \}&=  \{ I_{(+)}, H'_{(-)} \} = 0
    \end{split}\\[1.4ex]
\label{eq:rigidsusylast}
    \{ H_{(\pm)}, G' \} &= \{ I_{(\pm)}, G' \} = 0
  \end{align}
From \eqref{eq:susyham} we see that this algebra is of the type
\eqref{eq:admham}-\eqref{eq:virasoro2.2} rather than
\eqref{eq:psmham}-\eqref{eq:virasoro1.2}. 

Requirements (1) and (2) are fulfilled trivially, requirement (3) even holds
without the Lorentz constraint $G^\phi = G$ appearing on the right hand side
of \eqref{eq:rigidsusy2}.
Replacing $H_{(\pm)}$ by the alternative choice \eqref{eq:psmham} 
$\tilde{H}_{(+)} = H_{(+)} + G^\phi p_\phi$ and $\tilde{H}_{(-)} = H_{(-)}$
we arrive at
\begin{align}
  \{ I_{(+)}, I'_{(+)} \} &=  - 2  (\tilde{H}_{(+)} - G^\phi p_\phi)
  \delta(x-x')\ , \medsp  \{ I_{(-)}, I'_{(-)} \} &=  2  \tilde{H}_{(-)}
  \delta(x-x')\ .
\end{align}
Thus in contrast to \eqref{eq:rigidsusy1}-\eqref{eq:rigidsusylast} the algebra
of rigid supersymmetry involving $\tilde{H}_{(\pm)}$ is no longer linear.
\subsection{Algebra for generic gPSMs}
\label{sec:threetwo}
We now extend the considerations of the last subsection to general supergravity
models. Before evaluating the ``dilaton deformed'' algebra we have to define a
suitable extension of \eqref{eq:susyham} (or its counterpart
$\tilde{H}_{(\pm)}$) and \eqref{eq:rigidI}. Here the second
requirement of the introduction is essential: $H_{(\pm)}$ should retain the
structure as given in \eqref{eq:susyham}, whereas for the  the fermionic
generators the form \eqref{eq:rigidI} implies
\begin{align}
\label{eq:generalI}
    I_{(+)} &= \pm a G^+ \sqrt{p_{++}}\ , & I_{(-)} &= \pm a G^-
    \sqrt{p_{--}}\ .
\end{align}
Of course, for the most general form of $I_{(\pm)}$ the ansatz $I_{(\pm)} = \pm a
G^{\pm} \sqrt{p_{\pm\pm}} + \tilde{I}_{(\pm)}$ 
could be made. $\tilde{I}_{(\pm)}$ would be an \emph{arbitrary} fermionic
operator built from the variables in
\eqref{eq:canonicalalg}-\eqref{eq:canonicalalgII}. The limit of rigid
supersymmetry would dictate $\tilde{I}_{(\pm)} \Rightarrow 0$. Now requirement
(2) simply states that $\tilde{I}_{(\pm)} \equiv
0$ has been chosen. An analogous argument from requirement (2) precludes a
further fermionic extension of $H_{(\pm)}$. Requirement (1) suggests $a =
\mbox{const.}$ in \eqref{eq:generalI}, but we will have to check the
consequences of $a(\phi,Y)$ as well later on.

With the definitions
\eqref{eq:susyham}, resp.\ \eqref{eq:generalI} the general superconformal
algebra, after a laborious but straightforward calculation, using
\eqref{eq:canonicalalg}-\eqref{eq:canonicalalgII} becomes
  \begin{align}
    \begin{split}
\label{eq:gensupconf}
      \{ H_{(+)}, H'_{(+)} \} &=  (H_{(+)} + H'_{(+)}) \partial
    \delta(x-x')\ , \medsp
      \{ H_{(-)}, H'_{(-)} \} &=  - (H_{(-)} + H'_{(-)}) \partial
    \delta(x-x')\ , \medsp
      \{ H_{(+)}, H'_{(-)} \} &= G \partial_\phi ( P^{-- K} p_K p_{--} + P^{- K} p_K
    p_-) \delta(x-x')\ ,
    \end{split}\\[1.4ex]
    \begin{split}
\label{eq:gensupconf2}
      \{ I_{(+)}, I'_{(+)} \} &=  - a^2 (G^+ \partial_{++} P^{+ K} p_K + G^K
      \partial_K P^{+| +}p_{++})   \delta(x-x')\ , \medsp
      \{ I_{(-)}, I'_{(-)} \} &=  - a^2 (G^- \partial_{--} P^{- K} p_K + G^K
      \partial_K P^{-| -}p_{--})   \delta(x-x')\ , \medsp
      \{ I_{(+)}, I'_{(-)} \} &= - a^2 \Bigl(G^+ \partial_{++} P^{- K} p_K
      \frac{\sqrt{p_{--}}}{2 \sqrt{p_{++}}} + G^- \partial_{--} P^{+ K} p_K
      \frac{\sqrt{p_{++}}}{2 \sqrt{p_{--}}} \medsp
      &\quad \phantom{- a^2 \Bigl(} + G^K \partial_K P^{+-} \sqrt{p_{++}}
      \sqrt{p_{--}} \Bigr) \delta(x-x')\ ,
    \end{split}\\[1.4ex]
    \begin{split}
\label{eq:gensupconf3}
    \{ I_{(+)}, H'_{(+)} \} &=  (I_{(+)} + \half{1} I'_{(+)}) \partial
    \delta(x-x')\medsp
    &\quad - (\pm \frac{a}{2}) ( G \partial_\phi P^{+ K} p_K
    \sqrt{p_{++}} - \mathcal{D}^+) \delta(x-x')\ ,\medsp
   \{ I_{(-)}, H'_{(-)} \} &=  - (I_{(-)} + \half{1} I'_{(-)}) \partial
    \delta(x-x')\medsp
    &\quad + (\pm \frac{a}{2}) ( G \partial_\phi P^{- K} p_K
    \sqrt{p_{--}} + \mathcal{D}^-) \delta(x-x')\ ,\medsp
    \{ I_{(-)}, H'_{(+)} \}&= - (\pm \frac{a}{2}) ( G \partial_\phi P^{- K} p_K
    \sqrt{p_{--}} - \mathcal{D}^-) \delta(x-x')\ ,\medsp
    \{ I_{(+)}, H'_{(-)} \} &= (\pm \frac{a}{2}) ( G \partial_\phi P^{+ K} p_K
    \sqrt{p_{++}} + \mathcal{D}^+) \delta(x-x')\ ,
    \end{split}\\[1.4ex]
\label{eq:gensupconflast}
    \{ H_{(\pm)}, G' \} &= \{ I_{(\pm)}, G' \} = 0\ ,
  \end{align}
where $\mathcal{D}^\pm$ in $\{ I_{(\pm)}, H'_{(\pm)} \}$ are abbreviations for
  \begin{align}
    \begin{split}
      \label{eq:Dcalplus}
      \mathcal{D}^+ &= - \half{1} G^+ \partial_{++} \bigl( P^{++ K} p_K
      \sqrt{p_{++}} - P^{-- K} \frac{p_K
      p_{--}}{\sqrt{p_{++}}} + P^{+ K} \frac{ p_K p_+}{\sqrt{p_{++}}} -  P^{- K}
      \frac{ p_K p_-}{\sqrt{p_{++}}} \bigr) \medsp
      &\quad - G^K \partial_K \bigl(P^{+ | ++} p_{++} - P^{+| --} p_{--} + P^{+| +}
      p_{+} - P^{+| -} p_{-}\bigr) \sqrt{p_{++}} \medsp
      &\quad + \bigl(G^{++} \partial_{++} - G^{--} \partial_{--} + G^+ \partial_+ - G^-
      \partial_-\bigr) P^{+K} p_K \sqrt{p_{++}}\ ,
    \end{split}\medsp
    \begin{split}
      \label{eq:Dcalminus}
            \mathcal{D}^- &= - \half{1} G^- \partial_{--} \bigl( P^{++ K}
      \frac{p_K p_{++}}{
      \sqrt{p_{--}}} - P^{-- K} p_K \sqrt{p_{--}} + P^{+ K} \frac{ p_K p_+}{\sqrt{p_{--}}} -  P^{- K}
      \frac{ p_K p_-}{\sqrt{p_{--}}} \bigr) \medsp
      &\quad - G^K \partial_K \bigl(P^{-| ++} p_{++} - P^{-| --} p_{--} + P^{-| +}
      p_{+} - P^{-| -} p_{-}\bigr) \sqrt{p_{--}} \medsp
      &\quad + \bigl(G^{++} \partial_{++} - G^{--} \partial_{--} + G^+ \partial_+ - G^-
      \partial_-\bigr) P^{-K} p_K \sqrt{p_{--}}\ .
    \end{split}
  \end{align}
The components of the Poisson tensor $P^{IJ}$ in
\eqref{eq:gensupconf}-\eqref{eq:gensupconflast} are still completely general,
except for those dictated by local Lorentz transformations (cf.\  eqs.\ \eqref{eq:rigidgPSM}
and \eqref{eq:lorentzconstr}). Comparing with \eqref{eq:rigidsusy1}-\eqref{eq:rigidsusylast} we note that the
deformation of the algebra does not lead to additional structure functions
$\propto \partial \delta$, which happens to be a consequence of our choice of
requirement (2). Obviously in supergravity no simple redefinition
of the constraints can lead to an algebra including structure \emph{constants} only
instead of structure \emph{functions}. The main obstacle
is the fact that any redefinition of the constraints $H_{(\pm)}$, simplifying the purely
bosonic part of the algebra, inevitably leads to new non-linear terms in the
anti-commutator of two spinorial generators.

So far $a$ in equation \eqref{eq:generalI} has been assumed to be constant. That this
is in agreement with requirement (1) can be seen from the origin of that
constant. In the general
decomposition\footnote{For full fermionic rank, i.e.\ $\det v^{\alpha \beta}
  \neq 0$ only $\gamma_c$ and $\gthree$ are allowed (cf.\ after eq.\
  \eqref{eq:gammadef}). The Majorana
  nature of $\chi$ forbids terms $\mathcal{O}(\chi^3)$.} of $P^{\alpha \beta}$
\begin{equation}
\label{eq:palhphabetadecomp}
      P^{\alpha \beta} = v^{\alpha \beta} + \half{1} \chi^2 v_2^{\alpha
      \beta}\ ,
\end{equation}
in the term proportional to $\gamma_c$ in
\begin{align}
\label{eq:generalutilde}
      v^{\alpha \beta} &= i \tilde{u} X^c \gamma_c^{\alpha \beta} + u
      \gthree^{\alpha \beta} \ ,
\end{align}
requirement (1) at the gPSM level, comparing with the case of rigid
supersymmetry \eqref{eq:rigidgPSM2}, suggests
to take $\tilde{u} = \tilde{u}_0 =  (\sqrt{2} a^2)^{-1}$ to be
constant. Indeed $a$ so far has been chosen in that way.

One should, nevertheless, check the consequences of
relaxing this restriction and allowing $a$ in \eqref{eq:generalI} to become a function of the dilaton field
$\phi$ and of $Y = X^a X_a /2$. We first treat $a = a(\phi)$. Then additional
deformations emerge in certain commutators among \eqref{eq:gensupconf}-\eqref{eq:gensupconflast}:
\begin{equation}
\label{eq:phideform}
\begin{split}
  \{ I_{(+)}, I'_{(-)} \} &= (\pm) \partial_\phi a \bigl( P^{- \phi} \sqrt{p_{--}} I_{(+)}
  - P^{+ \phi} \sqrt{p_{++}} I_{(-)} \bigr) \delta(x-x') +\ldots \medsp
     \{ I_{(+)}, H'_{(+)} \} &= - (\pm) \partial_\phi \ln a \bigl( P^{++ \phi} p_{++} + P^{+
     \phi} p_+ \bigr) I_{(+)} \delta(x-x') + \ldots \medsp
   \{ I_{(+)}, H'_{(-)} \} &=
     (\pm) \partial_\phi \ln a \bigl( P^{-- \phi} p_{--} + P^{-
     \phi} p_- \bigr) I_{(+)} \delta(x-x') + \ldots \medsp
     \{ I_{(-)}, H'_{(+)} \} &= - (\pm) \partial_\phi \ln a \bigl( P^{++ \phi} p_{++} + P^{+
     \phi} p_+ \bigr) I_{(-)} \delta(x-x') + \ldots \medsp
   \{ I_{(-)}, H'_{(-)} \} &=
     (\pm) \partial_\phi \ln a \bigl( P^{-- \phi} p_{--} + P^{-
     \phi} p_- \bigr) I_{(-)} \delta(x-x') + \ldots
\end{split}
\end{equation}
The commutators $\{ I_{(+)}, I'_{(+)} \}$ and $\{ I_{(-)}, I'_{(-)} \}$ do not
experience any change. If $a$ becomes a function of $Y$, $\{ I_{(+)}, I'_{(+)} \}$ receives
additional contributions
\begin{equation}
\label{eq:Ydeform}
  \{ I_{(+)}, I'_{(+)} \} = -  a P^{+ c} \partial_c a G^+ p_{++} \delta(x - x')\ .
\end{equation}
The right hand side of \eqref{eq:Ydeform}
cannot be made a contribution to the Hamiltonian \eqref{eq:susyham} as
dictated for this anti-commutator by requirement (3). The only other
possibility would be $P^{+c} = 0$. But this restriction would lead back to the
special case of rigid supersymmetry\footnote{A different Poisson tensor
  (``block diagonal supergravity'') with $P^{+c} = 0$ had been presented in
  \cite{Ertl:2000si}. But in that case the most general $P^{\alpha \beta}$ has
the form $P^{\alpha \beta} = u(\phi,Y) (\gthree)^{\alpha \beta}$, precluding
the interpretation of the fermionic $\epsilon_\alpha$ as supersymmetry
transformations.}. Thus $ \tilde{u} = \tilde{u}(Y)$ for any non-trivial
algebra would always contradict requirement (3). If $\tilde{u} =
  \tilde{u}(\phi,Y)$ both deformations \eqref{eq:phideform} and
  \eqref{eq:Ydeform} add. Thus from now on only a dependence $\tilde{u} =
  \tilde{u}(\phi)$ will be permitted, because an $Y$-dependence has been excluded.

The absence of the Lorentz constraint $G$ on the r.h.s\
  of $\{ I_{(+)}, I'_{(+)} \}$ and $\{ I_{(-)}, I'_{(-)} \}$ for general
$\tilde{u} = \tilde{u}(\phi,Y)$ implies that the third
requirement of section \ref{sec:introd} can be tightened for the choice of
  $H_{(\pm)}$ as in \eqref{eq:susyham}:
\begin{enumerate}
\item[(3a)] The constraints $I_{(+)}$ with itself and $I_{(-)}$ with itself  anti-commute
back into \( H_{\left( 0\right) } \) and \( H_{\left( 1\right) } \).
\end{enumerate}

Still the general structure of \eqref{eq:gensupconf}-\eqref{eq:Dcalminus} shows the possibility of singularities at
$p_{++} = e_{1|++} = 0$ and $p_{--} = e_{1|--} = 0$ (cf.\ eq.\
\eqref{eq:gaugepot}). These singularities will turn out to disappear when the content of requirement (3)
of the introduction (resp.\ (3a) above) is taken into account in full
mathematical detail and not only qualitatively, as has been done so far.

One could be tempted to pursue the complete procedure of this section by trying to
simplify the algebra by the use of $\tilde{H}_{(\pm)}$ instead of
$H_{(\pm)}$. However, it can be seen easily that one would be forced to weaken the supersymmetry relation of the Hamiltonian constraints, being the
``square'' of the supersymmetry generators. It turns out that this does not
remove the complicated structure of $\{I_{(\pm)}, H'_{(\pm)} \}$, in particular it does
not remove the dependence of these commutators on the constraints $G^a$ and
$G^\alpha$.

It may be useful to summarize the results obtained so far: As a
consequence of requirements (1) and (2) the Hamiltonian
constraints are \eqref{eq:susyham}, while the fermionic ones must be chosen as in
\eqref{eq:generalI}, the latter being suggested by the
special case of rigid supersymmetry. The dependence of the factor $a = (\sqrt{2} \tilde{u})^{-1/2}$ on $Y$ is forbidden, but it
still may depend on $\phi$ at this point. Only an analytic implementation of
requirement (3) in section \ref{sec:genSUGRA} will show that, at the end of
the day a constant $a$, resp.\ $\tilde{u} = \tilde{u}_0$ alone will be allowed.


\section{Dilaton prepotential algebra: a special example}
\label{sec:dilatonprepot}
In order to check whether the three requirements are satisfied by \emph{some} gPSM
model with non-trivial gravity sector, we
illustrate the above algebra by an important
example, the dilaton prepotential algebra. This model originally had been
studied in ref.\ \cite{Izquierdo:1998hg} within a different approach, a derivation in terms of graded Poisson Sigma Models has been
given in ref.\ \cite{Ertl:2000si}. As the bosonic
potential is restricted here to be a function of the dilaton field $\phi$
only, these models have vanishing bosonic torsion ($Z = 0$ in
\eqref{eq:bosonpot}). For a Poisson tensor, which shares the first term in
\eqref{eq:dpptensorlast} with the one in rigid supersymmetry
\eqref{eq:rigidgPSM2}, the solution of \eqref{eq:nijenhuis} has been given in ref.\
\cite{Ertl:2000si}, eqs. (5.34)-(5.36) ($u' = \partial u/\partial \phi$)
  \begin{align}
\label{eq:dpptensor}
    P^{ab} &= \inv{2 \tilde{u}_0^2} \bigl( - (u^2)' + \half{1} \chi^2 u''
    \bigr) \epsilon^{ab}\ , \medsp
    P^{\alpha b} &= \frac{i u'}{2 \tilde{u}_0} (\chi \gamma^b)^\alpha\ , \medsp
\label{eq:dpptensorlast}
    P^{\alpha \beta} &= i \tilde{u}_0 X^c \gamma_c^{\alpha \beta} + u
    \gthree^{\alpha \beta}\ ,
  \end{align}
where $\tilde{u}_0 = \mbox{const.}$ and $u(\phi)$ is a ``prepotential'' related
to the bosonic potential $V$ in \eqref{eq:bosonpot} as
(cf. eq.\ (5.30) of ref.\ \cite{Ertl:2000si})
\begin{equation}
\label{eq:dppu}
  \tilde{u}_0^2 V + u u' = 0\ .
\end{equation}
In terms of \eqref{eq:rigidgPSM} and
\eqref{eq:dpptensor}-\eqref{eq:dpptensorlast} the gPSM action \eqref{eq:action}
\begin{multline}
  \label{eq:dppaction}
  \Stext{DPP} = \int_{\mathcal{M}} \bigl( \phi \diff \omega + X^a D e_a + \chi^\alpha D
  \psi_\alpha + \epsilon \Bigl( V - \inv{4u} \chi^2 \bigl( V' +
  4 \frac{V^2}{u^2} \bigr) \Bigr) \medsp
   + \frac{i V}{u}
  (\chi \gamma^a e_a \psi)
   + i X^a (\psi \gamma_a \psi) - \half{1} u (\psi \gamma_3 \psi) \bigr)
\end{multline}
with the covariant derivatives \eqref{eq:A8} includes the standard
contribution to the supertorsion from the fermions only, as can be observed
from the factor of $X^a$. This class of models already covers the
supersymmetrized version of several known two-dimensional gravity
models\footnote{The ``topological'' supergravity studied some time
ago \cite{Rivelles:1994xs,Leite:1995sa} also may be interpreted as a similar
gPSM with vanishing prepotential. On the other hand, it contains further
generators from a central extension of the algebra.}, in
particular \eqref{eq:SRG} and \eqref{eq:JT}. Eliminating \cite{Ertl:2000si} algebraically the auxiliary fields
$X^a$ and $\omega$ in \eqref{eq:dppaction} one arrives at a superdilaton
action of the form
\begin{multline}
\label{eq:simplifiedaction2}
  \Stext{DPP} = \intd{\diff{^2 x}} e \biggl( \half{1} \Tilde{R} \phi +
  (\chi \Tilde{\sigma}) + V - \inv{4 u} \chi^2 \bigl(V' + 4
  \frac{V^2}{u^2} \bigr)  \medsp
  - \frac{i V}{u} \epsilon^{mn} (\chi \gamma_n \psi_m) + \half{u}
  \epsilon^{mn} (\psi_n \gthree \psi_m) \biggr)\ .
\end{multline}
The Ricci scalar $\tilde{R} = 2 \ast \extd \tilde{\omega}$ is expressed in
terms of the dependent spin connection
\begin{equation}
  \tilde{\omega}_a = \epsilon^{mn} \partial_n e_{ma} - i \epsilon^{mn} (\psi_n
  \gamma_a \psi_m)\ .
\end{equation}
Its fermionic counterpart is
\begin{equation}
    \tilde{\sigma}_\alpha = \ast (\tilde{D} \psi)_\alpha =
  \epsilon^{mn} \bigl( \partial_n \psi_{m\alpha} + \half{1} \tilde{\omega}_n (\gthree
  \psi_m)_\alpha \bigr)\ .
\end{equation}
$V$ and $u$ are still related by \eqref{eq:dppu}.
This new form of the action turns out to be a special case of the superdilaton
action obtained from superspace as introduced by Park and Strominger \cite{Park:1993sd}, corresponding to $J(\Phi) = \Phi$ and
$K(\Phi) = 0$ in the notation of ref.\ \cite{Park:1993sd}. The relation between the
dilaton supergravity by Park and Strominger and this special case following
from a gPSM has been observed in ref.\ \cite{Ikeda:1994dr}, a more detailed
discussion of this topic will be given in \cite{Bergamin:2002}. In addition
this model can be related to the supergravity model of Howe quite
generally \cite{Ertl:2000si,Bergamin:2002}.

Together with the
relations \eqref{eq:lorentzconstr} the constraint algebra \eqref{eq:constralg}
for \eqref{eq:dpptensor}-\eqref{eq:dpptensorlast} becomes
  \begin{align}
    \{ G^a, G^b \} &= - \frac{\epsilon^{ab}}{2 \tilde{u}_0^2} \bigl( G^\phi(
    -(u^2)'' + \half{1} \chi^2 u''') + G^\alpha \chi_\alpha u''\bigr)\ , \medsp
    \{ G^\alpha, G^b \} &= - \frac{i}{2\tilde{u}_0} \bigl( G^\phi u''
    (\chi \gamma^b)^\alpha + G^\beta {(\gamma^b)_\beta}^\alpha \bigr)\ , \medsp
    \{ G^\alpha, G^\beta \} &= -i \tilde{u}_0 G^a \gamma_a^{\alpha \beta} -
    G^\phi u' \gthree^{\alpha \beta}\ .
  \end{align}
After a straightforward calculation the corresponding superconformal algebra is obtained in
terms of $H_{(\pm)}$ and $I_{(\pm)}$, which according to requirement (2) are
given by eqs.\ \eqref{eq:susyham} and \eqref{eq:generalI}: 
  \begin{align}
    \begin{split}
\label{eq:dppalg}
            \{ H_{(+)}, H'_{(+)} \} &=  (H_{(+)} + H'_{(+)}) \partial
    \delta(x-x') \\
      \{ H_{(-)}, H'_{(-)} \} &=  - (H_{(-)} + H'_{(-)}) \partial
    \delta(x-x') \\
      \{ H_{(+)}, H'_{(-)} \} &= G  \bigl(\inv{2 \tilde{u}_0^2} ( (u^2)''+ q^+ q^-
    u''') p_{++} p_{--}\\
    &\quad \phantom{G \bigl(}- \frac{u''}{\sqrt{2}
    \tilde{u}_0} (q^- p_+ p_{--} + q^+ p_- p_{++}) - u' p_+ p_-\bigr)
    \delta(x-x')
    \end{split}
\end{align}
\begin{align}
    \begin{split}
      \{ I_{(+)}, I'_{(+)} \} &=  - 2  H_{(+)}  \delta(x-x') \\
      \{ I_{(-)}, I'_{(-)} \} &=  2  H_{(-)}  \delta(x-x') \\
      \{ I_{(+)}, I'_{(-)} \} &= \pm \frac{\sqrt{2} i}{\tilde{u}_0} G u' \sqrt{p_{++}}
      \sqrt{p_{--}} \delta(x-x')
    \end{split}\\[1.4ex]
    \begin{split}
\label{eq:dppalglast}
          \{ I_{(+)}, H'_{(+)} \} &=  (I_{(+)} + \half{1} I'_{(+)}) \partial
    \delta(x-x')\\
   \{ I_{(-)}, H'_{(-)} \} &=  - (I_{(-)} + \half{1} I'_{(-)}) \partial
    \delta(x-x') \\
    \{ I_{(-)}, H'_{(+)} \}&= \pm \sqrt{\frac{\sqrt{2}}{\tilde{u}_0}} G
    (\frac{u''}{\sqrt{2}\tilde{u}_0} q^+ p_{++} + u' p_+ ) \sqrt{p_{--}}
    \delta(x-x')\medsp
    \{ I_{(+)}, H'_{(-)} \} &= \pm \sqrt{\frac{\sqrt{2}}{\tilde{u}_0}} G
    (\frac{u''}{\sqrt{2}\tilde{u}_0} q^- p_{--} + u' p_- ) \sqrt{p_{++}}
    \delta(x-x') \delta(x-x')
    \end{split}
  \end{align}
Evidently the deformation is restricted to the appearance
of the Lorentz constraint G which only occurs in the ``mixed''
(anti-)commutators of  $\left\{ H_{\left( +\right)}, H_{\left( -\right)
    }\right\}$, $\left\{ I_{\left( +\right) },I_{\left( -\right) }\right\}$,
  $\left\{ I_{\left( -\right) },H_{\left( +\right) }\right\}$ and
\( \left\{ I_{\left( +\right) },H_{\left( -\right) }\right\}  \).
Clearly the algebra obeys all the requirements set out in section
\ref{sec:introd}. Thus the
identification of this model with others obtained from standard supergravity requirements
\cite{Freedman:1976xh,Freedman:1976py,Deser:1976eh,Deser:1976rb,Grimm:1978kp}
imply an important a posteriori confirmation of our requirements, as far as they have been
taken into account already. However, our requirements will not only allow this special
case. They will identify a more general class of gPSM derived theories with $Z
\neq 0$ in \eqref{eq:bosonpot}, whose relation to models obtained from
superspace has not been demonstrated so far in the literature.  

Writing down the algebra in terms
$\tilde{H}_{(\pm)}$ of eqs.\ \eqref{eq:psmham} does not provide any new insight into the structure of this
theory. In contrast to the bosonic case \eqref{eq:virasoro1.1} the undeformed
superconformal algebra is found for
constant $u$ only, which corresponds to rigid supersymmetry.

In view of the singularity problems of a generic gPSM \cite{Ertl:2000si}
it is important to study the regularity of the algebra about which
no assumptions have been made so far. An obvious constraint is $\tilde{u}_0
\neq 0$, which however is
irrelevant from the physical point of view: For $\tilde{u}_0 = 0$
supersymmetry transformations are no longer generated
by  $G^\alpha$ and thus the meaning of
the model is lost.  Furthermore in contrast to the general algebra
\eqref{eq:gensupconf}-\eqref{eq:gensupconflast}, the dilaton prepotential
supergravity is regular for $p_a = 0$, which, in this particular case, is easily seen to be a trivial
consequence of  vanishing bosonic torsion: all potentially singular terms in
eqs.\ \eqref{eq:gensupconf2}, \eqref{eq:Dcalplus} and \eqref{eq:Dcalminus} are
multiplied by vanishing derivatives $\partial_c P^{\alpha a}$ and $\partial_c
P^{+ | -}$.


\section{General deformed superconformal algebra}
\label{sec:genSUGRA}
We now show that the most general supergravity model
described by a gPSM, obeying the restrictions of section \ref{sec:introd},
must exhibit a similar structure as the
dilaton prepotential algebra of section \ref{sec:dilatonprepot}. The only possible generalization will turn out
to be closely
related to the algebra \eqref{eq:dppalg}-\eqref{eq:dppalglast} as well,
because it may be produced 
by a special target space diffeomorphism from the latter.

The requirements (1) and (2) are implemented
already in \eqref{eq:gensupconf}-\eqref{eq:gensupconflast}.
Starting from the most general gPSM algebra
\eqref{eq:gensupconf}-\eqref{eq:gensupconflast} we now demand the anti-commutators
$\{ I_{(+)}, I'_{(+)}\}$ and $\{ I_{(-)}, I'_{(-)} \}$ to yield \emph{exactly} the Hamiltonian
$H_{(\pm)}$ (requirement (3a)).

Comparison of the
right hand side of the first two equations of
\eqref{eq:gensupconf2}\footnote{The comment immediately after eq.\
  \eqref{eq:phideform} is important here.} with the
desired expression for $H_{(\pm)}$ (cf.\ \eqref{eq:susyham}) translates  into
six necessary and sufficient conditions on the Poisson tensor:
\begin{equation}
\label{eq:regcond}
\begin{split}
  \derfrac{X^{++}}(P^{+|-}, P^{+|++}, P^{+| --}) &= 0 \\
  \derfrac{X^{--}} ( P^{+|-}, P^{-|--},  P^{-| ++}) &= 0
\end{split}
\end{equation}
As an example we consider the r.h.s.\ of the first equation of
\eqref{eq:gensupconf2}. Due to the restrictions that followed from the
discussion in section \ref{sec:threetwo}, the second term of that expression
can contribute with an expression of the form $G^{++} p_{++}$ only, which is
part of $H_{(+)}$. The first term in the first eq.\ of
\eqref{eq:gensupconf2}, however, contributes  $G^+ p_{++}$, $G^+
p_{--}$ and $G^+ p_-$, which do not appear in $H_{(+)}$. They are eliminated
by the first three conditions in \eqref{eq:regcond}.

In the following we need the general explicit solution of the vanishing graded Nijenhuis
tensor \eqref{eq:nijenhuis} already obtained in ref.\ \cite{Ertl:2000si}. As
we do not want to repeat the lengthy calculation leading to this result, we
refer the reader for all notations and details
of the calculation to this work, especially sections 3, 5.4 and 5.5. If the
technical details of the detailed implementation of
\eqref{eq:regcond} are not interesting to the reader, he/she may jump to the
final result as
given in \eqref{eq:mostgensup}-\eqref{eq:mostgensuplast} together with
\eqref{eq:finalpot} below. Within the given approach this represents the most
general Poisson tensor describing supergravity.

The argument starts with the case $\tilde{u} = \tilde{u}_0 = \mbox{constant}$. Applying \eqref{eq:regcond} to the general decomposition for $P^{\alpha
  \beta}$ of eqs.\  \eqref{eq:palhphabetadecomp} and \eqref{eq:generalutilde} restricts its terms as
\begin{align}
\label{eq:constutilde}
      v^{\alpha \beta} &= i \tilde{u}_0 X^c \gamma_c^{\alpha \beta} + u(\phi)
      \gthree^{\alpha \beta}\ , & v_2^{\alpha \beta} &= \tilde{v}(\phi)
      \gthree^{\alpha \beta}\ .
    \end{align}
The bosonic part $P^{ab}$ is still general (cf.\ \cite{Ertl:2000si})
\begin{equation}
\label{eq:pabdecomp}
  P^{ab} = \epsilon^{ab} \bigl( v(\phi,Y) + \half{1} \chi^2 v_2(\phi,Y)\bigr)\ .
\end{equation}
The first term in \eqref{eq:pabdecomp} is the input from the bosonic theory \eqref{eq:bosonicPSM},
\eqref{eq:bosonpot}, the second is the only possible one in presence of the dilatino. As shown in ref.\ \cite{Ertl:2000si} the complete
solution from the vanishing Nijenhuis tensor is parametrized by $v^{\alpha
  \beta}$, $v$ and a
Lorentz vector field $f^a$, which appears in
the decomposition of the ``mixed'' component of the Poisson tensor 
\begin{equation}
\label{eq:Falphabdecomp}
  P^{\alpha b} = \chi^\beta {{F^b}_\beta}^\alpha\ .
\end{equation}
Together with other terms it creates a contribution to the decomposition
  \eqref{eq:Falphabdecomp} of the form
  \begin{equation}
\label{eq:unwantedFc}
    {{F^c}_\alpha}^\beta = i f(\phi) X^c X^b {\gamma_{b \alpha}}^\beta +
    \ldots\ ,
  \end{equation}
with an invariant function $f(\phi)$. In ref.\
\cite{Ertl:2000si} in order to eliminate the term \eqref{eq:unwantedFc}  the
special choice $f^a = \half{1} u' X^a$ has been made. In that reference this was
motivated by the desire to present a simple solution and, at the same time, to
make contact with an earlier result of ref.\ \cite{Izquierdo:1998hg}. In our
present context a term like \eqref{eq:unwantedFc}, producing in
\eqref{eq:Falphabdecomp} contributions to $P^{\alpha a}$ of type
\begin{align}
  P^{+| ++} &\sim X^{++}
X^{++} \chi^- + \ldots\ , & P^{+| --} &\sim X^{--}
X^{++} \chi^- + \ldots
\end{align}
would be in obvious contradiction to \eqref{eq:regcond}. Thus the \emph{same}
choice of $f^a$ as in ref.\ \cite{Ertl:2000si} must be made. It should be noted that the
remaining terms in ${{F^c}_\alpha}^\beta$ yield $P^{+| ++} \equiv 0$.

Due to the restrictions from \eqref{eq:regcond} found so far according to
\eqref{eq:constutilde} and \eqref{eq:unwantedFc}, 
$v_2^{\alpha \beta}$ in \eqref{eq:constutilde} and
$v_2$ in \eqref{eq:pabdecomp} can be parametrized in terms of two remaining
functions of the dilaton field: the bosonic potential $v(\phi)$ and the
prepotential $u(\phi)$. As this specific solution has been selected for a quite \emph{different} reason
already in ref.\ \cite{Ertl:2000si} (cf.\ above) we are allowed to take over the final
result from there (eqs.\ (5.27)-(5.29) in \cite{Ertl:2000si}):                    
  \begin{align}
\label{eq:fcalphabeta}
    {{F^c}_\alpha}^\beta &= \inv{2 \Delta} ( \tilde{u}_0^2 v + u u') X^a
    {(\gamma_a \gamma^c \gthree)_\alpha}^\beta + \frac{i \tilde{u}_0}{2
    \Delta} ( uv + 2 Y u') { \gamma^c_\alpha}^\beta\ , \medsp
\label{eq:v2alphabeta}
  v_2^{\alpha \beta} &= \inv{2 \Delta} ( \tilde{u}_0^2 v + u u')
    \gthree^{\alpha \beta}\ , \medsp
\label{eq:v2}
    v_2 &= \frac{uv}{2 \Delta^2} ( \tilde{u}_0^2 v + u u') + \frac{u u'}{2
    \Delta^2} ( uv + 2 Y u') + \inv{2 \Delta} (u v' + 2 Y u' \dot{v} + 2 Y
    u'')\ ,
  \end{align}
with
\begin{equation}
\label{eq:Delta}
\Delta = 2 Y \tilde{u}_0^2 - u^2
\end{equation}
being the determinant of $v^{\alpha
  \beta}$. Derivatives with respect to the dilaton field are indicated by a
  prime, derivatives with respect to $Y$ by a dot.

So far the function $u$ in the solution and the bosonic potential $v$ were not
related. This residual
freedom can now be used to achieve the cancellation of the $X^a$-dependence as
required by \eqref{eq:regcond} also for the components of the Poisson tensor still
generated by \eqref{eq:fcalphabeta} and \eqref{eq:v2alphabeta}. Starting from
\eqref{eq:v2alphabeta}, $\partial_{++} P^{+|-} = 0$ and noting that $X^a$ only
appears in the $Y$-dependence of $v$ and $\Delta$ implies that the factor of
$\gthree$ in \eqref{eq:v2alphabeta} must be a function of $\phi$
alone. Therefore, a potential linear in $Y$ of \eqref{eq:bosonpot} turns out
to be the only allowed one! The ensuing relation 
\begin{equation}
\label{eq:finalpot}
V\left( \phi \right) =-\frac{\left( u^{2}\right)'
}{2\tilde{u}_0^2}-\frac{u^{2}}{2\tilde{u}_0^{2}}Z\left( \phi
\right)\ ,
\end{equation}
with the definition of $\Delta$ may be rewritten as  
\begin{equation}
\label{eq:torsionpot}
  v = - \frac{(u^2)'}{2 \tilde{u}_0} + \inv{2 \tilde{u}_0^2} Z(\phi) \Delta \ .
\end{equation}
However, adopting \eqref{eq:torsionpot} for $v$ is found to eliminate at
the same time \emph{all} singularities $\propto
\Delta^{-1}$ also in \eqref{eq:fcalphabeta} and \eqref{eq:v2}. The only
remaining condition from \eqref{eq:regcond}, $\partial_{++} P^{+|--} = 0$, is
fulfilled automatically.

The divergence
problem of \eqref{eq:fcalphabeta}-\eqref{eq:v2} at $\Delta = 0$ had been discussed
extensively in ref.\ \cite{Ertl:2000si}. Therein the
cancellations of these divergences was introduced as an \emph{ad hoc assumption}
and the most general function $v$ leading to that cancellation was found to be
\begin{equation}
  v = - \frac{(u^2)'}{2 \tilde{u}_0} + T(\Delta, \phi, Y)
\end{equation}
with $\frac{ T(\Delta, \phi, Y)}{\Delta}$ regular at $\Delta = 0$. Eq.\
\eqref{eq:torsionpot}, obtained here by a different route, is a special case of
that.

The fact that the new poles at $\Delta = 0$ vanish at the level of the Poisson tensor is
another confirmation of the appropriateness of our conditions (1) to (3) of
section \ref{sec:introd} (resp. (3a) of section \ref{sec:gengPSM}). However, at first
sight singularities at $p_{\pm \pm} = 0$  seem to persist at the level of the
algebra \eqref{eq:gensupconf}-\eqref{eq:gensupconflast}. Inspecting the
structure of the remaining deformations in the general
algebra, it turns out that they disappear if besides \eqref{eq:regcond} in addition
\begin{align}
\label{eq:suffcondition}
  \partial_{++} P^{++|--} &= 2 \partial_+ P^{+| --}\ , & \partial_{++} P^{++|-}
  &= 2 \partial_+ P^{+|-}
\end{align}
holds, a condition which is indeed met in all models allowed by our requirements. To
see this eq.\ \eqref{eq:torsionpot} is used in
\eqref{eq:fcalphabeta}-\eqref{eq:v2} to eliminate $\Delta$. Then the the
components \eqref{eq:palhphabetadecomp}, \eqref{eq:pabdecomp} and
\eqref{eq:Falphabdecomp} of the Poisson tensor become
\begin{align}
\label{eq:mostgensup}
  P^{ab} &= \biggl( V + Y Z - \half{1} \chi^2 \Bigl( \frac{VZ + V'}{2u} +
  \frac{\tilde{u}_0^2 V^2}{2 u^3} \Bigr) \biggr) \epsilon^{ab}\ , \medsp
  P^{\alpha b} &= \frac{Z}{4} X^a
    {(\chi \gamma_a \gamma^b \gthree)}^\alpha - \frac{i \tilde{u}_0 V}{2 u}
  (\chi \gamma^b)^\alpha\ , \medsp
\label{eq:mostgensuplast}
  P^{\alpha \beta} &= i \tilde{u}_0 X^c \gamma_c^{\alpha \beta} + \bigl( u +
  \frac{Z}{8} \chi^2 \bigr) \gthree^{\alpha \beta}\ .
\end{align}
It can be checked that the factors in front of the expressions involving
$Z$ exactly guarantee that the conditions \eqref{eq:suffcondition} are
satisfied. We conclude that our original conditions together with the
symmetries of the Poisson tensor were also sufficient to ensure a regular
superconformal algebra at $p_{\pm \pm} = 0$.

At the level of the models discussed so far ($\tilde{u} = \tilde{u}_0 = \mbox{constant}$) we
observe that the following three conditions are equivalent:
\renewcommand{\labelenumi}{(\Alph{enumi})}
\begin{enumerate}
\item $\{ I_+, I'_+ \}$ and $\{ I_-, I'_- \}$ have no
  deformations, which agrees with requirement (3) of section \ref{sec:introd}.
\item All deformations with respect to a superconformal algebra depend on the Lorentz constraint only.
\item The fermionic extensions of the Poisson tensor are regular at $\Delta = 0$ and the algebra is regular
  at $p_{\pm \pm} = 0$.
\end{enumerate}
Condition (B) obviously implies condition (A), but
it is a remarkable result that our original condition is actually
sufficient. We have shown above that condition (A) implies
condition (C). To see that the reverse is true as well it is
important to notice that we have to cancel the $f_{(t)}$ term in
\eqref{eq:unwantedFc}, although $P^{+|++}$ now must be independent of $X^{--}$
and not of $X^{++}$.

A final argument is required with respect to a dependence $\tilde{u}_0 \rightarrow
\tilde{u}(\phi)$ in \eqref{eq:constutilde}, which is still an open possibility. This type of
algebra has been studied in ref.\ \cite{Ertl:2000si} as well. Choosing again the
independent vector field $f^a$ to obtain a cancellation of \eqref{eq:unwantedFc},
\eqref{eq:fcalphabeta} in that case generalizes to
($\nabla = \partial_\phi - v \partial_Y$)
\begin{equation}
  F^c = - \inv{4} (\nabla \ln \Delta)  X^a
    (\gamma_a \gamma^c \gthree) + \half{1}(\nabla \ln \tilde{u}) X^c \gthree +
    \frac{i \tilde{u} u}{2 \Delta} \Bigl( v + 2 Y \bigl( \nabla \ln
    \frac{u}{\tilde{u}}\bigr) \Bigr) \gamma^c \ .
\end{equation}
Here both $u$ (note that the argument leading from \eqref{eq:palhphabetadecomp} to
\eqref{eq:constutilde} is still valid) and $\tilde{u}$
(cf.\ \eqref{eq:Ydeform}) are still allowed to depend on
the dilaton field only. But when  $\partial_\phi \tilde{u}
\neq 0$ the term $\propto \gthree$ does not vanish and this leads to $X^c$-dependent
contributions to $P^{+| ++}$ and $P^{-| --}$ which contradict \eqref{eq:regcond}. It should be noted that they are
never able to
cancel together with the other contributions to this part of the tensor, which
we have set to zero by means of the independent vector field $f^a$. Thus
only $\tilde{u} = \tilde{u}_0 = \mbox{const.}$ is allowed by \eqref{eq:regcond} and 
\eqref{eq:mostgensup}-\eqref{eq:mostgensuplast} is indeed the most general
model obeying condition (A), resp.\ (3a).

Although (A) and (B) are still equivalent there exist
some models with $\tilde{u} = \tilde{u}(\phi)$ that obey the 
condition (C) without fulfilling (A) and (B). An example of this type has been given in ref.\
\cite{Ertl:2000si}, where (in
section 5.7.1) a supersymmetric extension of spherically reduced gravity not
fulfilling (A) and (B) can be found.

A more general $\tilde{u}(\phi, Y)$ already has been excluded in section
\ref{sec:threetwo} (cf.\ \eqref{eq:Ydeform}). It is  remarkable that a non-constant $\tilde{u} = \tilde{u}(\phi,Y)$ is not
compatible too with the constraints \eqref{eq:regcond} on the Poisson
tensor. Indeed $\tilde{u}(Y)$ independent of the dilaton field is
excluded as well, as $v \equiv 0$ is not a solution in this case. It remains
the possibility of $\tilde{u}(\phi, Y)$ with $\nabla \tilde{u} = 0$. To cancel
all $Y$-dependent terms in $F^c$ which disagrees with \eqref{eq:regcond} we would have to choose $v = -2 Y
\frac{u'}{u}$. But with that $v$  one finds that $v_2^{\alpha \beta}$
cannot be independent of $Y$ as well.

Starting from a given bosonic potential \eqref{eq:bosonpot}, the main
restriction to the application of a ``genuine'' supergravity extension turns
out to be \eqref{eq:finalpot}. It is important to note that    
spherically reduced gravity \eqref{eq:SRG} fits into this structure as do the
models \eqref{eq:JT} and \eqref{eq:howe}, but a
model quadratic in torsion and curvature
\cite{Katanaev:1986wk,Katanaev:1990qm} for certain values
of its parameters does not in general, or, at best, experiences restrictions
on the range of its bosonic fields.

As shown in ref.\ \cite{Ertl:2000si} the class of models determined by
\eqref{eq:torsionpot}, resp.\ \eqref{eq:finalpot} can be obtained also by a
conformal transformation of a dilaton prepotential model of section
\ref{sec:dilatonprepot}, which may be interpreted as a special target space diffeomorphism
of a gPSM. All this is a strong indication that among the gPSM models the ones
with tensor \eqref{eq:mostgensup}-\eqref{eq:mostgensuplast} and the
prepotential determined by the solution of \eqref{eq:finalpot} represent the most
general ``genuine'' $N = (1,1)$ supergravity in $d = 2$. The ensuing disappearance of the
singularities in $\Delta$ and $p_{\pm\pm}$ in our approach is a fortuitous but welcome
``accident''. According to the solution of a general gPSM in
\cite{Ertl:2000si} not obeying our requirements, generic singularities at $Y =
0$ may be expected as well. Also those are absent here.
\section{General deformed supergravity}
\label{sec:finalalg}
Having shown that the most general supergravity model is the
supersymmetrization of a bosonic model with a potential $V$ linear in $Y$
(cf.\ \eqref{eq:bosonpot}, but note the prepotential type restriction \eqref{eq:finalpot} on $V$!) its
(deformed) superalgebra can now be given explicitly. We economize writing it
down in full detail by
the observation  that it shares most of its structure with the one of the dilaton prepotential algebra \eqref{eq:dppalg}-\eqref{eq:dppalglast}:
\begin{itemize}
\item The commutators of the form $\{ A_{(+)}, B_{(+)}\}$ and $\{ A_{(-)},
  B_{(-)}\}$  exactly reproduce the non-trivial part of the well-known
  superconformal algebra. All deformations are in the commutators of the form
  $\{ A_{(+)}, B_{(-)}\}$. This property is not difficult to read off from
  \eqref{eq:gensupconf}-\eqref{eq:gensupconflast}: $\{I_{(+)}, I'_{(+)}\}$ and
  $\{I_{(-)}, I'_{(-)}\}$ have been
  arranged such that they precisely yield $H_{(+)}$ and $H_{(-)}$. For $\{ I_{(+)},
  H'_{(+)} \}$ and $\{ I_{(-)},
  H'_{(-)} \}$ one simply observes that
  the conditions imposed on the Poisson tensor directly lead to the last two
  equations of \eqref{eq:gensupconf3} with the much simpler quantities
  \begin{equation}
\begin{split}
    \mathcal{D}^+ &= G \partial_\phi (P^{+| --} p_{--} + P^{+|-} p_-)
    \sqrt{p_{++}}\ , \medsp
    \mathcal{D}^- &=-  G \partial_\phi (P^{-| ++} p_{++} + P^{+|-} p_+)
    \sqrt{p_{--}}\ .
\end{split}
  \end{equation}
\item The deformation in the remaining commutators is a slight generalization
  of the dilaton prepotential algebra of section \ref{sec:dilatonprepot} ($a =
  (\sqrt{2} \tilde{u}_0)^{-1/2}$):
    \begin{align}
\begin{split}
      \{ H_+, H'_- \} &= G \partial_\phi ( P^{--|++} p_{++} p_{--} + P^{--|
      +} p_{+} p_{--}\\
    &\quad \phantom{ G \partial_\phi (}+ P^{-|++} p_{++} p_{-} + P^{+|-} p_{+} p_{-} )
      \delta(x-x')
\end{split}\\
      \{ I_+, I'_- \} &= - a^2 G \partial_\phi P^{+|-} \sqrt{p_{++}} \sqrt{p_{--}}
      \delta(x-x') \\
      \{ I_+, H'_- \} &= (\pm a) \mathcal{D}^+ \delta(x-x')\\
      \{ I_-, H'_+ \} &= (\pm a) \mathcal{D}^- \delta(x-x')
    \end{align}
\end{itemize}
This similar structure, of course, is not just a coincidence but is contingent
upon the fact that the models of section
\ref{sec:dilatonprepot} and the most general one, obeying our requirements (1),
(2)
and (3) (resp.\ (3a)), are related by a simple conformal transformation (cf.\
ref.\ \cite{Ertl:2000si}, section 5.5).

For completeness we present the action of the most general
``genuine'' supergravity model as obtained on the basis of the gPSM formalism. Inserting the corresponding Poisson tensor
\eqref{eq:mostgensup}-\eqref{eq:mostgensuplast} into eq.\ \eqref{eq:action}
leads to
\begin{multline}
  \label{eq:mostgenaction}
  \mathcal{S} = \int_{\mathcal{M}} \bigl( \phi \diff \omega + X^a D e_a + \chi^\alpha D
  \psi_\alpha + \epsilon \biggl( V + Y Z - \half{1} \chi^2 \Bigl( \frac{VZ + V'}{2u} +
  \frac{\tilde{u}_0^2 V^2}{2 u^3} \Bigr) \biggr) \medsp
   + \frac{Z}{4} X^a
    (\chi \gamma_a \gamma^b e_b \gthree \psi) - \frac{i \tilde{u}_0 V}{2 u}
  (\chi \gamma^a e_a \psi) \medsp
   - \frac{i \tilde{u}_0}{2} X^a (\psi \gamma_a \psi) - \half{1} \bigl( u +
  \frac{Z}{8} \chi^2 \bigr) (\psi \gamma_3 \psi) \bigr)
\end{multline}
Starting from a bosonic gravity theory constructed according to the principles
outlined in section \ref{sec:introd} and equipped with a potential as given in
\eqref{eq:bosonpot}, by our analysis this is its unique $N = (1,1)$
supergravity extension and it contains the minimal number of fields: The
bosonic variables $( \omega, e_a)$ and $(\phi, X^a)$ are supplemented by a
gravitino $\psi_\alpha = \psi_{\alpha m} \extd x^m$ and a dilatino
$\chi^\alpha$, only. No auxiliary fields appear, nevertheless
\eqref{eq:mostgenaction} has a (relatively) simple structure.


\section{Conclusions}
\label{sec:conclusions}
The underlying symmetries in the PSM-formulation produce a simple,
albeit nonlinear algebra of Hamiltonian constraints which is difficult to analyze
from the point of view of supersymmetry and supergravity. However,
as noticed some time ago for 2d bosonic gravity theories \cite{Katanaev:1994qf}
by a linear transformation of the constraints this algebra can be cast into the one appearing
for the constraints in the ADM-analysis (``Virasoro type algebra''), supplemented by the Lorentz
constraint in a simple manner, when
the deformation by the dilaton field becomes relevant. In two dimensions
such an algebra in the parlance of the string community is also
called a ``conformal algebra''. Here we have used this method
for the graded case. 

A natural set of requirements for a ``genuine'' supergravity theory
(``superconformal gravity'') implies that rigid supersymmetry
appears in some flat limit and that some obvious main features of
the superconformal algebra should survive in the deformed one. Starting from the requirement (1) of section \ref{sec:introd} (rigid supersymmetry in the flat
limit), its implementation together with (2) determines the proper choice for
the constraints
$H_{(\pm)}$ and $I_{(\pm)}$ to be used in the dilaton supergravity
algebra. While (1) and (2) do not yet lead to restrictions on the relation
between the bosonic
potential $v$ and the prepotential $u$, requirement (3) of sect.\ \ref{sec:introd}, resp.\ (3a) of sect.\ \ref{sec:gengPSM}, ($ \{ I_{(+)}, I'_{(+)} \} =  - 2  H_{(+)}
\delta(x-x')$, etc.)\ leads to the mathematical condition
\eqref{eq:regcond}. Its application drastically reduces the class of allowed Poisson tensors
and exhibits very gratifying properties:

\begin{itemize}
\item gPSM supergravity leads to a \emph{unique} fermionic extension of all
  bosonic PSM theories of gravity, whose potential, however, is forced to be,
  at most, quadratic in torsion (cf.\ \eqref{eq:mostgenaction}). Moreover the
  bosonic potential must be derivable from a prepotential (cf.\ \eqref{eq:finalpot}).
\item The supergravity algebra of these general supergravities does not show the
  singularities and obstructions, which in the the generic case of gPSMs are
  present.
\item The most general supergravity action allowed
  by the requirements does not
  lead to \emph{any} new singularities in the fermionic extension.
\end{itemize}

The bosonic models
which possess this unique supergravity extension are precisely the one
which include spherically reduced gravity, the (bosonic) stringy Black
Hole and other physically motivated theories. It should be emphasized that
none of the final ``fortious'' results regarding uniqueness and the absence of singularities and
obstructions were
introduced in an ad-hoc manner.

The present argument has been applied to the $N=(1,1)$ supergravities
of refs.\ \cite{Ertl:2000si,Ertl:2001sj}. Spherical reduction of $N=1$ supergravity in $d=4$
is known to need Dirac Killing spinors. Therefore, a treatment of
that case has to await the extension of the gPSM approach to $N=(2,2)$
which should be possible along the the same lines. It would be especially
interesting to check also in that case the consequences of the three requirements
imposed in our present work.

Other promising directions of research seem to be the investigation
of global properties by means of a super pointparticle for the class
of ``physical'' 2d supergravity theories determined here and the coupling of
matter fields. The superparticle in supergravity so far has been formulated in
the superfield formalism only. To this end the relation of gPSM supergravity to superspace
supergravity should be clarified. So far, this has been done only partially
\cite{Ertl:2001sj,Ikeda:1994fh,Ikeda:1994dr}, namely for models with $Z = 0$
in \eqref{eq:bosonpot}, related to a non-dynamical dilaton field in the
equivalent action \eqref{eq:GDT}. Recent calculations
for \emph{non-vanishing} bosonic torsion ($Z \neq 0$) indicate that there exists a close relation between precisely the gPSM supergravities
found in this work and the dilaton supergravity formulated in superspace
\cite{Bergamin:2002}. As the determination of general classical
solutions and the quantization in the PSM formulation is a well-established
field \cite{Grumiller:2002nm,Bergamin:2001}, this opens new perspectives for dilaton \emph{superfield}
supergravity \cite{Park:1993sd}, problems for which complete solutions
involving the fermionic sector do not seem to exist in the literature.

\subsection*{Acknowledgement}
The authors are grateful especially to D. Grumiller for
stimulating discussions which led us to consider the approach of this
paper. They thank T.\ Strobl for illuminating comments. The work has been supported by the Swiss National Science Foundation (SNF) and the Austrian Science Foundation
(FWF) Project 14650-TPH and P-16030-N08.


\appendix
\section{Notations and conventions}
In this technical Appendix notations and conventions are
summarized briefly. They are identical to those in ref.\
\cite{Ertl:2000si,Ertl:2001sj}, where additional explanations can be found.

A rather elaborate labelling of indices is used to distinguish different
combinations of objects:
\begin{description}
\item[$\mathbf{I, J, K}$] Capital Latin letters from the middle of the alphabet represent a
  generic index, which can include both, commuting and anti-commuting
  objects. Commutations of objects are defined in the standard way
  \begin{equation}
    v^I w^J = (-1)^{IJ} w^J v^I\ ,
  \end{equation}
  where the indices in the exponent take values 0 (commuting object) or 1
  (anti-commuting object). In the application to gPSM supergravity the generic
  index $I$ itself splits into the classes
  \begin{equation}
    I = (i,\alpha) = (\phi, a, \alpha)\ ,
  \end{equation}
  which are explained below.
\item[$\mathbf{i, j, k}$] Lower case Latin letters from the middle of the alphabet represent a
  generic commuting object.
\item[$\boldsymbol{\phi}$] The index $\phi$ is used to label the dilaton component of
  the gPSM fields:
  \begin{align}
    X^\phi &= \phi & A_\phi &= \omega
  \end{align}
\item[$\mathbf{a, b, c}$] Lower case Latin letters from the beginning of the
  alphabet are used as bosonic anholonomic indices. At the level of gPSM
  supergravity they label the two bosonic one-forms representing the zweibein
  and the corresponding target-space variables, resp.
\item[$\mathbf{\boldsymbol{\alpha}, \boldsymbol{\beta}, \boldsymbol{\gamma}}$] Greek letters from the beginning of the alphabet are
  (anholonomic) anti-commuting indices. They label the spinor components of the
  gravitino and the dilatino.
\item[$\mathbf{m, n, o}$] To label the components of the world-sheet manifold lower
  case Latin letters from the middle of the alphabet are used.
\end{description}

The summation convention is always $NW \rightarrow SE$, especially for a
fermion $\chi$: $\chi^2 = \chi^\alpha \chi_\alpha$. Our conventions are
arranged in such a way that almost every bosonic expression is transformed
trivially to the graded case when using this summation convention and
replacing commuting indices by general ones. This is possible together with
exterior derivatives acting \emph{from the right}, only. Thus the graded
Leibniz rule is given by
\begin{equation}
  \label{eq:leibniz}
  \mbox{d}\left( AB\right) =A\mbox{d}B+\left( -1\right) ^{B}(\mbox{d}A) B\ .
\end{equation}

In terms of anholonomic indices the metric and the symplectic $2 \times 2$
tensor are defined as
\begin{align}
  \eta_{ab} &= \eta^{ab} = \left( \begin{array}{cc} 1 & 0 \\ 0 & -1 \end{array} \right) &
  \epsilon_{ab} &= - \epsilon^{ab} = \left( \begin{array}{cc} 0 & 1 \\ -1 & 0
  \end{array} \right) & \epsilon_{\alpha \beta} &= \epsilon^{\alpha \beta} = \left( \begin{array}{cc} 0 & 1 \\ -1 & 0
  \end{array} \right)
\end{align}
The metric in terms of holonomic indices is obtained by $g_{mn} = e_n^b e_m^a
\eta_{ab}$ and for the determinant the standard expression $e = \det e_m^a =
\sqrt{- \det g_{mn}}$ is used. The volume form reads $\epsilon = \half{1}
\epsilon^{ab} e_b \wedge e_a$.

The $\gamma$-matrices are used in a chiral representation:
\begin{align}
\label{eq:gammadef}
  {{\gamma^0}_\alpha}^\beta &= \left( \begin{array}{cc} 0 & 1 \\ 1 & 0
  \end{array} \right) & {{\gamma^1}_\alpha}^\beta &= \left( \begin{array}{cc} 0 & 1 \\ -1 & 0
  \end{array} \right) & {{\gthree}_\alpha}^\beta &= {(\gamma^1
    \gamma^0)_\alpha}^\beta = \left( \begin{array}{cc} 1 & 0 \\ 0 & -1
  \end{array} \right)
\end{align}
The matrices $(\gamma^a)^{\alpha \beta} = \epsilon^{\alpha \delta}
{{\gamma^a}_\delta}^\beta$ and $(\gthree)^{\alpha \beta}$ are symmetric in $\{
\alpha, \beta \}$
The most important relations among the $\gamma$-matrices are:
\begin{align}
  \gamma^a \gamma^b &= \eta^{ab} \mathbf{1} + \epsilon^{ab} \gthree &
  \gamma^a \gthree &= \gamma^b {\epsilon_b}^a
\end{align}

Covariant derivatives of anholonomic indices with respect to the geometry
include the two-dimensional spin-connection $\omega^{ab} = \omega
\epsilon^{ab}$. When acting on lower indices the explicit expressions read
($\half{1} \gthree$ is the generator of Lorentz transformations in spinor space):
\begin{align}
\label{eq:A8}
  (D e)_a &= \extd e_a - \omega {\epsilon_a}^b e_b & (D \psi)_\alpha &= \extd
  \psi_\alpha + \half{1} {{\omega \gthree}_\alpha}^\beta \psi_\beta
\end{align}

Finally light-cone components are introduced. As we work with spinors in a
chiral representation we can use
\begin{align}
  \chi^\alpha &= ( \chi^+, \chi^-)\ , & \chi_\alpha &= \begin{pmatrix} \chi_+ \\
  \chi_- \end{pmatrix}\ .
\end{align}
For Majorana spinors upper and lower chiral components are related by $\chi^+
= \chi_-$, $ \chi^- = - \chi_+$. Vectors in light-cone coordinates are given by
\begin{align}
\label{eq:A10}
  v^{++} &= \frac{i}{\sqrt{2}} (v^0 + v^1)\ , & v^{--} &= \frac{-i}{\sqrt{2}}
  (v^0 - v^1)\ .
\end{align}
Correspondingly the derivatives with respect to these components are
written compactly as
\begin{equation}
  \partial_K = \derfrac{X^K} = (\partial_\phi, \partial_{++}, \partial_{--},
  \partial_{+}, \partial_{-})\ .
\end{equation}
The additional factor $i$ in \eqref{eq:A10} permits a direct identification of the light-cone components with
the components of the spin-tensor $v^{\alpha \beta} = \frac{i}{\sqrt{2}} v^c \gamma_c^{\alpha
  \beta}$. This implies that $\eta_{++|--} = 1$
and $\epsilon_{--|++} = - \epsilon_{++|--} = 1$. The
$\gamma$-matrices in light-cone coordinates become
\begin{align}
\label{eq:gammalc}
  {(\gamma^{++})_\alpha}^\beta &= \sqrt{2} i \left( \begin{array}{cc} 0 & 1 \\ 0 & 0
  \end{array} \right)\ , & {(\gamma^{--})_\alpha}^\beta &= - \sqrt{2} i \left( \begin{array}{cc} 0 & 0 \\ 1 & 0
  \end{array} \right)\ .
\end{align}
To prevent confusion of indices in tensors we separate them by a vertical line
where necessary (e.g.\ $P^{++|+}$ is the component $a = ++$, $\alpha = +$ of
$P^{a \alpha}$).

\bibliography{biblio}

\providecommand{\href}[2]{#2}\begingroup\raggedright\begin{thebibliography}{10}

\bibitem{Dicke:1957}
R.~H. Dicke {\em Rev. Mod. Phys.} {\bf 29} (1957) 29.

\bibitem{jordanschwerkraft}
P.~Jordan, {\em {S}chwerkraft und {W}eltall : {G}rundlagen der theoretischen
  {K}osmologie}.
\newblock Vieweg, second~ed., 1955.

\bibitem{Jordan:1959eg}
P.~Jordan, {\it The present state of {D}irac's cosmological hypothesis},  {\em
  Z. Phys.} {\bf 157} (1959) 112--121.

\bibitem{Fierz:1956}
M.~Fierz, {\it {\"U}ber die physikalische {D}eutung der erweiterten
  {G}ravitationstheorie {P.} {J}ordans},  {\em Helv. Phys. Acta} {\bf 29}
  (1956) 128.

\bibitem{Brans:1961sx}
C.~Brans and R.~H. Dicke, {\it Mach's principle and a relativistic theory of
  gravitation},  {\em Phys. Rev.} {\bf 124} (1961) 925--935.

\bibitem{Callan:1992rs}
C.~G. Callan, Jr., S.~B. Giddings, J.~A. Harvey, and A.~Strominger, {\it
  Evanescent black holes},  {\em Phys. Rev.} {\bf D45} (1992) 1005--1009,
  [\href{http://xxx.lanl.gov/abs/hep-th/9111056}{{\tt hep-th/9111056}}].

\bibitem{Grumiller:2002nm}
D.~Grumiller, W.~Kummer, and D.~V. Vassilevich, {\it Dilaton gravity in two
  dimensions},  {\em Phys. Rept.} {\bf 369} (2002) 327,
  [\href{http://xxx.lanl.gov/abs/http://arXiv.org/abs/hep-th/0204253}{{\tt
  http://arXiv.org/abs/hep-th/0204253}}].

\bibitem{Freedman:1976xh}
D.~Z. Freedman, P.~van Nieuwenhuizen, and S.~Ferrara, {\it Progress toward a
  theory of supergravity},  {\em Phys. Rev.} {\bf D13} (1976) 3214--3218.

\bibitem{Freedman:1976py}
D.~Z. Freedman and P.~van Nieuwenhuizen, {\it Properties of supergravity
  theory},  {\em Phys. Rev.} {\bf D14} (1976) 912.

\bibitem{Deser:1976eh}
S.~Deser and B.~Zumino, {\it Consistent supergravity},  {\em Phys. Lett.} {\bf
  B62} (1976) 335.

\bibitem{Deser:1976rb}
S.~Deser and B.~Zumino, {\it A complete action for the spinning string},  {\em
  Phys. Lett.} {\bf B65} (1976) 369--373.

\bibitem{Green:1987sp}
M.~B. Green, J.~H. Schwarz, and E.~Witten, {\em SUPERSTRING THEORY}.
\newblock Cambridge University Press, 1987.
\newblock Vol. 1: {I}ntroduction.

\bibitem{lust89}
D.~L{\"u}st and S.~Theisen, {\em Lectures on String Theory}, vol.~346 of {\em
  Lecture notes in physics}.
\newblock Springer Verlag, Berlin, 1989.

\bibitem{Polchinski:1998rq}
J.~Polchinski, {\em String theory}.
\newblock Cambridge University Press, 1998.
\newblock Vol. I and II.

\bibitem{Haider:1994cw}
F.~Haider and W.~Kummer, {\it {Quantum functional integration of nonEinsteinian
  gravity in d = 2}},  {\em Int. J. Mod. Phys.} {\bf A9} (1994) 207--220.

\bibitem{Kummer:1997hy}
W.~Kummer, H.~Liebl, and D.~V. Vassilevich, {\it Exact path integral
  quantization of generic 2-d dilaton gravity},  {\em Nucl. Phys.} {\bf B493}
  (1997) 491--502,
  [\href{http://xxx.lanl.gov/abs/http://arXiv.org/abs/gr-qc/9612012}{{\tt
  http://arXiv.org/abs/gr-qc/9612012}}].

\bibitem{Kummer:1998jj}
W.~Kummer, H.~Liebl, and D.~V. Vassilevich, {\it Non-perturbative path integral
  of 2d dilaton gravity and two-loop effects from scalar matter},  {\em Nucl.
  Phys.} {\bf B513} (1998) 723--734,
  [\href{http://xxx.lanl.gov/abs/http://arXiv.org/abs/hep-th/9707115}{{\tt
  http://arXiv.org/abs/hep-th/9707115}}].

\bibitem{Kummer:1998zs}
W.~Kummer, H.~Liebl, and D.~V. Vassilevich, {\it Integrating geometry in
  general 2d dilaton gravity with matter},  {\em Nucl. Phys.} {\bf B544} (1999)
  403--431, [\href{http://xxx.lanl.gov/abs/hep-th/9809168}{{\tt
  hep-th/9809168}}].

\bibitem{Katanaev:1996bh}
M.~O. Katanaev, W.~Kummer, and H.~Liebl, {\it {Geometric Interpretation and
  Classification of Global Solutions in Generalized Dilaton Gravity}},  {\em
  Phys. Rev.} {\bf D53} (1996) 5609--5618,
  [\href{http://xxx.lanl.gov/abs/http://arXiv.org/abs/gr-qc/9511009}{{\tt
  http://arXiv.org/abs/gr-qc/9511009}}].

\bibitem{Katanaev:1997ni}
M.~O. Katanaev, W.~Kummer, and H.~Liebl, {\it On the completeness of the black
  hole singularity in 2d dilaton theories},  {\em Nucl. Phys.} {\bf B486}
  (1997) 353--370, [\href{http://xxx.lanl.gov/abs/gr-qc/9602040}{{\tt
  gr-qc/9602040}}].

\bibitem{Mann:1990gh}
R.~B. Mann, A.~Shiekh, and L.~Tarasov, {\it Classical and quantum properties of
  two-dimensional black holes},  {\em Nucl. Phys.} {\bf B341} (1990) 134--154.

\bibitem{Banks:1991mk}
T.~Banks and M.~O'Loughlin, {\it Two-dimensional quantum gravity in {M}inkowski
  space},  {\em Nucl. Phys.} {\bf B362} (1991) 649--664.

\bibitem{Gegenberg:1993rg}
J.~Gegenberg and G.~Kunstatter, {\it Quantum theory of black holes},  {\em
  Phys. Rev.} {\bf D47} (1993) 4192--4195,
  [\href{http://xxx.lanl.gov/abs/gr-qc/9302006}{{\tt gr-qc/9302006}}].

\bibitem{Lemos:1994py}
J.~P.~S. Lemos and P.~M. Sa, {\it The black holes of a general two-dimensional
  dilaton gravity theory},  {\em Phys. Rev.} {\bf D49} (1994) 2897--2908,
  [\href{http://xxx.lanl.gov/abs/arXiv:gr-qc/9311008}{{\tt
  arXiv:gr-qc/9311008}}].

\bibitem{Louis-Martinez:1994eh}
D.~Louis-Martinez, J.~Gegenberg, and G.~Kunstatter, {\it Exact {D}irac
  quantization of all 2-d dilaton gravity theories},  {\em Phys. Lett.} {\bf
  B321} (1994) 193--198, [\href{http://xxx.lanl.gov/abs/gr-qc/9309018}{{\tt
  gr-qc/9309018}}].

\bibitem{Schaller:1994np}
P.~Schaller and T.~Strobl, {\it {Canonical quantization of non-Einsteinian
  gravity and the problem of time}},  {\em Class. Quant. Grav.} {\bf 11} (1994)
  331--346, [\href{http://xxx.lanl.gov/abs/arXiv:hep-th/9211054}{{\tt
  arXiv:hep-th/9211054}}].

\bibitem{Schaller:1994es}
P.~Schaller and T.~Strobl, {\it Poisson structure induced (topological) field
  theories},  {\em Mod. Phys. Lett.} {\bf A9} (1994) 3129--3136,
  [\href{http://xxx.lanl.gov/abs/http://arXiv.org/abs/hep-th/9405110}{{\tt
  http://arXiv.org/abs/hep-th/9405110}}].

\bibitem{Thomi:1984na}
P.~Thomi, B.~Isaak, and P.~H{\'a}j{\'i}{\v{c}}ek, {\it Spherically symmetric
  systems of fields and black holes. 1. {D}efinition and properties of apparent
  horizon},  {\em Phys. Rev.} {\bf D30} (1984) 1168.

\bibitem{Hajicek:1984mz}
P.~H{\'a}j{\'i}{\v{c}}ek, {\it Spherically symmetric systems of fields and
  black holes. 2. {A}pparent horizon in canonical formalism},  {\em Phys. Rev.}
  {\bf D30} (1984) 1178.

\bibitem{Schmidt:1997mq}
H.-J. Schmidt, {\it A new proof of birkhoff's theorem},  {\em Grav. Cosmol.}
  {\bf 3} (1997) 185,
  [\href{http://xxx.lanl.gov/abs/http://arXiv.org/abs/gr-qc/9709071}{{\tt
  http://arXiv.org/abs/gr-qc/9709071}}].

\bibitem{Schmidt:1998ih}
H.-J. Schmidt, {\it A two-dimensional representation of four-dimensional
  gravitational waves},  {\em Int. J. Mod. Phys.} {\bf D7} (1998) 215--224,
  [\href{http://xxx.lanl.gov/abs/http://arXiv.org/abs/gr-qc/9712034}{{\tt
  http://arXiv.org/abs/gr-qc/9712034}}].

\bibitem{Barbashov:1979bm}
B.~M. Barbashov, V.~V. Nesterenko, and A.~M. Chervyakov, {\it The solitons in
  some geometrical field theories},  {\em Theor. Math. Phys.} {\bf 40} (1979)
  572--581. Teor. Mat. Fiz. 40 (1979) 15--27, J. Phys. A13 (1980) 301--312.

\bibitem{Teitelboim:1983ux}
C.~Teitelboim, {\it Gravitation and {H}amiltonian structure in two space-time
  dimensions},  {\em Phys. Lett.} {\bf B126} (1983) 41.

\bibitem{teitelboim:1984}
C.~Teitelboim, {\it The hamiltonian structure of two-dimensional space-time and
  its relation with the conformal anomaly},  in {\em Quantum theory of gravity
  : essays in honor of the 60th birthday of Bryce S.DeWitt} (S.~Christensen,
  ed.), (Bristol), pp.~327--344, Hilger, 1984.

\bibitem{jackiw:1984}
R.~Jackiw, {\it Liouville field theory: a two-dimensional model for gravity},
  in {\em Quantum theory of gravity : essays in honor of the 60th birthday of
  Bryce S.DeWitt} (S.~Christensen, ed.), (Bristol), pp.~327--344, Hilger, 1984.

\bibitem{Jackiw:1985je}
R.~Jackiw, {\it Lower dimensional gravity},  {\em Nucl. Phys.} {\bf B252}
  (1985) 343--356.

\bibitem{Howe:1979ia}
P.~S. Howe, {\it Super {W}eyl transformations in two-dimensions},  {\em J.
  Phys.} {\bf A12} (1979) 393--402.

\bibitem{Ertl:2000si}
M.~Ertl, W.~Kummer, and T.~Strobl, {\it General two-dimensional supergravity
  from {P}oisson superalgebras},  {\em JHEP} {\bf 01} (2001) 042,
  [\href{http://xxx.lanl.gov/abs/arXiv:hep-th/0012219}{{\tt
  arXiv:hep-th/0012219}}].

\bibitem{Park:1993sd}
Y.-C. Park and A.~Strominger, {\it Supersymmetry and positive energy in
  classical and quantum two-dimensional dilaton gravity},  {\em Phys. Rev.}
  {\bf D47} (1993) 1569--1575,
  [\href{http://xxx.lanl.gov/abs/arXiv:hep-th/9210017}{{\tt
  arXiv:hep-th/9210017}}].

\bibitem{Ertl:2001sj}
M.~Ertl, {\em Supergravity in two spacetime dimensions}.
\newblock PhD thesis, {T}echnische {U}niversit{\"a}t {W}ien, 2001.
\newblock \href{http://xxx.lanl.gov/abs/arXiv:hep-th/0102140}{{\tt
  arXiv:hep-th/0102140}}.

\bibitem{Strobl:1999zz}
T.~Strobl, {\it Target-superspace in 2d dilatonic supergravity},  {\em Phys.
  Lett.} {\bf B460} (1999) 87--93,
  [\href{http://xxx.lanl.gov/abs/arXiv:hep-th/9906230}{{\tt
  arXiv:hep-th/9906230}}].

\bibitem{Grimm:1978kp}
R.~Grimm, J.~Wess, and B.~Zumino, {\it Consistency checks on the superspace
  formulation of supergravity},  {\em Phys. Lett.} {\bf B73} (1978) 415.

\bibitem{Grosse:1992vc}
H.~Grosse, W.~Kummer, P.~Presnajder, and D.~J. Schwarz, {\it {Novel symmetry of
  nonEinsteinian gravity in two- dimensions}},  {\em J. Math. Phys.} {\bf 33}
  (1992) 3892--3900, [\href{http://xxx.lanl.gov/abs/hep-th/9205071}{{\tt
  hep-th/9205071}}].

\bibitem{Katanaev:1994qf}
M.~O. Katanaev, {\it Canonical quantization of the string with dynamical
  geometry and anomaly free nontrivial string in two- dimensions},  {\em Nucl.
  Phys.} {\bf B416} (1994) 563--605,
  [\href{http://xxx.lanl.gov/abs/http://arXiv.org/abs/hep-th/0101168}{{\tt
  http://arXiv.org/abs/hep-th/0101168}}].

\bibitem{Katanaev:2000kc}
M.~O. Katanaev, {\it Effective action for scalar fields in two-dimensional
  gravity},  {\em Annals Phys.} {\bf 296} (2002) 1--50,
  [\href{http://xxx.lanl.gov/abs/http://arXiv.org/abs/gr-qc/0101033}{{\tt
  http://arXiv.org/abs/gr-qc/0101033}}].

\bibitem{Arnowitt:1962}
R.~Arnowitt, S.~Deser, and C.~W. Misner in {\em {Gravitation: An Introduction
  to Current Research}} (L.~Witten, ed.), Wiley, New York, 1962.

\bibitem{Izquierdo:1998hg}
J.~M. Izquierdo, {\it Free differential algebras and generic 2d dilatonic
  (super)gravities},  {\em Phys. Rev.} {\bf D59} (1999) 084017,
  [\href{http://xxx.lanl.gov/abs/arXiv:hep-th/9807007}{{\tt
  arXiv:hep-th/9807007}}].

\bibitem{Bergamin:2001}
L.~Bergamin, D.~Grumiller, W.~Kummer, and P.~van Nieuwenhuizen in preparation.

\bibitem{Rivelles:1994xs}
V.~O. Rivelles, {\it Topological two-dimensional dilaton supergravity},  {\em
  Phys. Lett.} {\bf B321} (1994) 189--192,
  [\href{http://xxx.lanl.gov/abs/http://arXiv.org/abs/hep-th/9301029}{{\tt
  http://arXiv.org/abs/hep-th/9301029}}].

\bibitem{Leite:1995sa}
M.~M. Leite and V.~O. Rivelles, {\it Topological dilatonic supergravity
  theories},  {\em Class. Quant. Grav.} {\bf 12} (1995) 627--636,
  [\href{http://xxx.lanl.gov/abs/http://arXiv.org/abs/hep-th/9410003}{{\tt
  http://arXiv.org/abs/hep-th/9410003}}].

\bibitem{Ikeda:1994dr}
N.~Ikeda, {\it Gauge theory based on nonlinear {L}ie superalgebras and
  structure of 2-d dilaton supergravity},  {\em Int. J. Mod. Phys.} {\bf A9}
  (1994) 1137--1152.

\bibitem{Bergamin:2002}
L.~Bergamin and W.~Kummer in preparation.

\bibitem{Katanaev:1986wk}
M.~O. Katanaev and I.~V. Volovich, {\it String model with dynamical geometry
  and torsion},  {\em Phys. Lett.} {\bf B175} (1986) 413--416.

\bibitem{Katanaev:1990qm}
M.~O. Katanaev and I.~V. Volovich, {\it Two-dimensional gravity with dynamical
  torsion and strings},  {\em Ann. Phys.} {\bf 197} (1990) 1.

\bibitem{Ikeda:1994fh}
N.~Ikeda, {\it Two-dimensional gravity and nonlinear gauge theory},  {\em Ann.
  Phys.} {\bf 235} (1994) 435--464,
  [\href{http://xxx.lanl.gov/abs/arXiv:hep-th/9312059}{{\tt
  arXiv:hep-th/9312059}}].

\end{thebibliography}\endgroup

\end{document}